\documentstyle[preprint,aps,epsf,floats]{revtex}

\tightenlines 

\newcommand{\beq}{\begin{equation}} \newcommand{\eeq}{\end{equation}}
\newcommand{\bqa}{\begin{eqnarray}} \newcommand{\eqa}{\end{eqnarray}}
\def\sumint{\hbox{$\sum$}\!\!\!\!\!\!\int}
\def\square{\vcenter{\vbox{\hrule height.4pt
          \hbox{\vrule width.4pt height8pt
          \kern8pt\vrule width.4pt}\hrule height.4pt}}}

\begin{document}
\vspace{-0.9cm}

\preprint{
\vbox{\halign{&##\hfil\cr
        & hep-ph/9905337  \cr
        & May 1999    \cr}}}

\title{ Hard-thermal-loop Resummation \\
                of the Thermodynamics of a Hot Gluon Plasma }

\author{Jens O. Andersen, Eric Braaten, and Michael Strickland}
\address{Physics Department, Ohio State University, Columbus OH 43210, USA}

\maketitle
\begin{abstract}
We calculate the thermodynamic functions
of a hot gluon plasma  to leading order in
hard-thermal-loop (HTL) perturbation theory.  Effects associated with
screening, gluon quasiparticles, and Landau damping are resummed
to all orders.  The ultraviolet divergences generated by the HTL
propagator corrections can be canceled by a 
counterterm that depends on the thermal gluon mass parameter.  
The HTL thermodynamic functions are compared to those from lattice gauge
theory calculations and from quasiparticle models. 
For reasonable values of the HTL parameters,
the deviations from lattice results
for $T>2T_c$
have the correct sign and roughly the correct magnitude to be
accounted for by next-to-leading order corrections in HTL perturbation theory.
\end{abstract}
\draft
\newpage
\section{Introduction}
Relativistic heavy-ion collisions will soon allow the experimental study of
hadronic matter at energy densities that will probably exceed 
that required to create a quark-gluon plasma.  
A quantitative understanding of the properties of a
quark-gluon plasma is essential in order to determine whether it has been
created.  Because QCD, the gauge theory that describes strong interactions,
is asymptotically free, its running coupling constant $\alpha_s$ becomes
weak at sufficiently high temperatures.  This would seem to make the task of
understanding the high-temperature limit of hadronic matter relatively
straightforward, because the problem can be attacked using perturbative
methods.  Unfortunately, a straightforward perturbative expansion in powers
of $\alpha_s$ does not seem to be of any quantitative use even at temperatures
that are orders of magnitude higher than those achievable in
heavy-ion collisions.

\begin{figure}[htb]


\hspace{1cm}
\epsfysize=10cm
\centerline{\epsffile{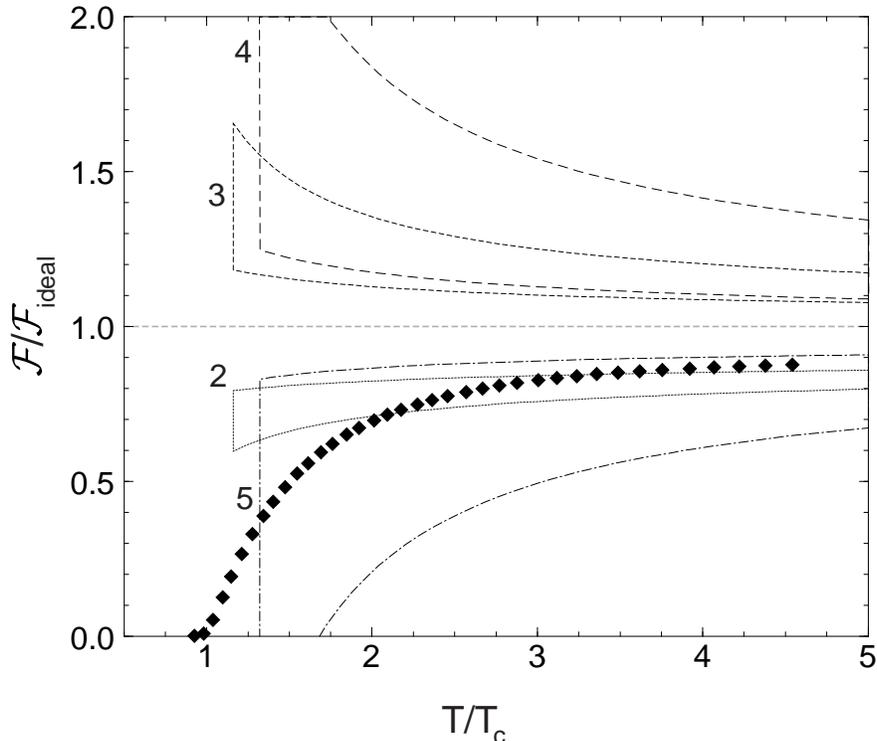}}
\caption[a]{The free energy for pure-glue QCD as a function of $T/T_c$.
The weak-coupling expansions through orders $\alpha_s$, $\alpha_s^{3/2}$, $\alpha_s^2$, and $\alpha_s^{5/2}$ are shown as bands that correspond to varying
the renormalization scale $\mu_4$ by a factor of two. 
The lattice results from Boyd et al.~\cite{lattice-0} are shown as diamonds.
}
\label{weakfig}
\end{figure}

The problem is evident in the free energy ${\cal F}$ of the quark-gluon
plasma, whose weak-coupling expansion has been calculated through order
$\alpha_s^{5/2}$ \cite{Kastening-Zhai,Kastening-Zhai2,Braaten-Nieto}.  An optimist might hope
to use perturbative methods at temperatures as low as 0.3 GeV, because the
running coupling constant $\alpha_s(2 \pi T)$ at the scale of the lowest
Matsubara frequency is about 1/3.  
Unfortunately, the weak-coupling expansion seems to diverge
badly even at much higher temperatures. 
For a pure-glue plasma, the first few terms in the weak-coupling expansion
are
\begin{eqnarray}\nonumber
{\cal F}_{\rm QCD} &=& {\cal F}_{\rm ideal} 
\Bigg[ 1 \;-\; {15 \over 4} {\alpha_s \over \pi} 
\;+\; 30 \left ( {\alpha_s \over \pi} \right )^{3/2} \\
&&\;+\; {135 \over 2} \left( \log {\alpha_s \over \pi} 
-{11\over36}\log{\mu_4\over2\pi T}+ 3.51 \right) 
	\left( {\alpha_s \over \pi} \right)^2 
\nonumber
\\
&&
+{495\over2}\left(\log{\mu_4\over2\pi T}-3.23\right) \left ( {\alpha_s \over \pi} \right )^{5/2}
\;+\; {\cal O} (\alpha_s^3\log \alpha_s ) \Bigg],
\label{F1-QCD}
\end{eqnarray}
where ${\cal F}_{\rm ideal}=-(8\pi^2/45)T^4$ is the free energy of an 
ideal gas of massless gluons and $\alpha_s=\alpha_s(\mu_4)$ is the running
coupling constant in the $\overline{\mbox{MS}}$ scheme.
In Fig.~\ref{weakfig}, the free energy is shown as a 
function of $T/T_c$, 
 where $T_c$ is the critical temperature
for the deconfinement transition.
The weak-coupling expansions through
orders $\alpha_s$, $\alpha_s^{3/2}$, $\alpha_s^2$, and $\alpha_s^{5/2}$
are shown as bands that correspond to varying the renormalization scale
$\mu_4$
by a factor of two from the central value $\mu_4=2\pi T$. 
As successive terms in the weak-coupling expansion
are added, the predictions fluctuate wildly and the sensitivity to the
renormalization scale grows.
Of course, because of asymptotic freedom,
the first few terms in the weak-coupling expansion will appear to 
converge at sufficiently high temperature. However, this occurs only at
temperatures orders of magnitude larger than $T_c$.
For example,
the $\alpha_s^{3/2}$ term is smaller than the 
$\alpha_s$ term only if $\alpha_s$ is less than about 1/20, 
which corresponds to a temperature
greater than about $10^5T_c$.
It is clear that a reorganization of the perturbative series 
is essential if perturbative calculations are to be of any quantitative
use at temperatures accessible in heavy-ion collisions.

The poor convergence of the perturbative series is puzzling, 
because lattice gauge theory calculations indicate that the 
free energy ${\cal F}$ of the quark-gluon plasma 
can be approximated by that of an ideal gas 
unless the temperature $T$ is very close to $T_c$~\cite{lattice-0,lattice-Nf}.
In Fig.~\ref{weakfig}, we also show the lattice results for the free energy
of pure-glue QCD from Boyd et al.~\cite{lattice-0}.
The free energy approaches that of an ideal gas of 
massless gluons from below as the temperature is increased and the deviation 
is less than about 25\% if $T$ is greater 
than $2 T_c$.

There are many possible ways
to reorganize the perturbation 
series for the free energy in order to improve its convergence.
One possibility is to apply Pad\'e approximation methods to the expansion
in $\alpha_s^{1/2}$~\cite{Pade}. This does 
improve the convergence somewhat, but it is at best a recipe with little 
physical motivation.
There are also technical problems associated with the fact that the
weak-coupling expansion is not simply a power series in $\alpha_s^{1/2}$.
Effective field theory methods can be used to show that it 
has the general form
\bqa
\label{fqcd}
{\cal F}_{\rm QCD}=T^4\sum_{n=0}^{\infty}c_n(\log\alpha_s)
\alpha_s^{n/2},
\eqa
where the coefficients $c_n(\log\alpha_s)$ are polynomials 
in $\log\alpha_s$~\cite{Braaten}.
The first few coefficients can be calculated using perturbative methods and
are given in~(\ref{F1-QCD}). Beginning at
order $\alpha_s^3$, nonperturbative methods are required to calculate some
of the coefficients.
Another problem with the Pad\'e approximation method is that it is
applicable only if several terms
in the perturbative series are known. This essentially limits its applicability
to the free energy ${\cal F}$ and to the other thermodynamic functions that
can be obtained from $\cal F$ by differentiation.
Almost all calculations involving signatures of the quark-gluon
plasma have been carried
out only to leading order. The only exception is the production of hard
dileptons which has been calculated to next-to-leading order in 
$\alpha_s$~\cite{ALT-AUR}. 

Another approach to the problem is based on the observation that the large 
corrections of order $\alpha_s^{3/2}$ and $\alpha_s^{5/2}$ in~(\ref{F1-QCD})
arise from the momentum scale $gT$. Effective field-theory
methods can be used to separate the momentum scale $T$ from the
lower scales $gT$ and $g^2T$. The resulting effective field theory,
dimensionally-reduced QCD (DRQCD), is an $SU(3)$
gauge theory with adjoint scalars in three euclidean dimensions.
The construction of DRQCD can be used to separate the free energy  
into contributions from the scale $T$
and from lower scales:
\bqa
\label{sepqcd}
{\cal F}(T)&=&{\cal F}_0(T,\alpha_s)+T
f_1(m_E^2,g_E^2,...).
\eqa
The function ${\cal F}_0$ and the parameters $m_E^2$, $g_E^2$, ...
of the effective field theory can be computed as power series in
$\alpha_s(2\pi T)$ by matching the effective field theory with thermal QCD.
For pure-glue QCD, the parameters of DRQCD at leading order
in $\alpha_s$ are $m_E^2=g^2T^2$, $g_E^2=g^2T$.
The function $f_1$ can be expanded in powers of $g_E^2/m_E$ and
the coefficients of the first three terms are known 
analytically~\cite{Braaten-Nieto}.
The first term gives the large $\alpha_s^{3/2}$ correction in~(\ref{F1-QCD}).
The third term contributes $-830(\alpha_s/\pi)^{5/2}$ to the
expansion in~(\ref{F1-QCD}) and therefore accounts for most of the
large $\alpha_s^{5/2}$ correction.
Thus the poor convergence properties
of the QCD perturbative series seems to arise
from a breakdown of perturbation theory in DRQCD.
This suggests that nonperturbative methods should be used to
compute the second term in~(\ref{sepqcd}) as a function of the
parameters $m_E^2$, $g_E^2$, ... of DRQCD. The only systematic nonperturbative
method currently available is Monte Carlo simulations of lattice
DRQCD. This method has been developed by Kajantie, Rummukainen, and
Shaposhnikov~\cite{KAJANTIE2}, and used to compute the Debye 
screening mass for thermal
QCD~\cite{debye}.
It has not yet been applied to the free energy. One disadvantage
of this method is that it relies in an essential way on the analytic 
continuation of the quantum field theory to imaginary time. It therefore cannot be used to
calculate signatures of the quark-gluon plasma that involve 
real-time processes.

The poor convergence of the weak-coupling expansion for the free energy
is not specific to QCD. The free energy of a massless scalar field
theory with a $\phi^4$ interaction
has also been computed to order $g^5$~\cite{Parwani}. 
The series seems to converge
only if the coupling constant $g$ is extremely small. 
The largest corrections, which include the $g^{3/2}$ term, come from
the momentum scale $gT$.
This suggests that
the source of the convergence problem could be similar for QCD and the 
massless scalar theory. 

There have been several other attempts to improve the convergence
of the free energy for the massless
scalar field theory~\cite{K-P-P,Rebhan}.  
One of the most successful approaches is ``screened
perturbation theory'' developed by Karsch, Patk\'os, and Petreczky\cite{K-P-P}.
This approach can be made more systematic by using the framework
of ``optimized perturbation theory'', which has been applied by
Chiku and Hatsuda to a scalar field theory with spontaneous symmetry
breaking~\cite{Chiku-Hatsuda}.
A local mass term proportional to $\phi^2$ is added and subtracted from
the lagrangian, with the added term included nonperturbatively and the
subtracted term treated as a perturbation.  
The renormalizability of the $\phi^2$ term
guarantees that all ultraviolet
divergences generated by the mass term can be systematically removed by
renormalization.
When the free energy is calculated using 
screened perturbation theory, the convergence of successive approximations 
to the free energy is dramatically improved.

Conventional perturbation theory is essentially an expansion about an
ideal gas of massless particles. 
Screened perturbation theory is 
a reorganization of perturbation
theory such that the expansion is 
about an ideal gas of quasiparticles with
a temperature-dependent mass.
Empirical  evidence that such a reorganization of perturbation theory
might be useful for QCD 
is provided by the success of quasiparticle models for the
thermodynamics of QCD at temperatures above $T_c$.
There were several early attempts to fit the thermodynamic functions
calculated by lattice gauge theory to those of an ideal gas of massive 
quarks and gluons~\cite{Biro+,Biro2}.
In recent years, there have been several significant improvements in
the lattice gauge theory calculations.
The thermodynamic functions for pure-glue QCD have been calculated with very
high precision by Boyd et al.~\cite{lattice-0}. There have also been 
calculations with $N_f=2$ and 4 flavors of dynamical quarks~\cite{lattice-Nf}.
Motivated by these developments, more quantitative comparisons with 
quasiparticle models have been carried out by Peshier et al.~\cite{quasi1}
and by L\'evai and Heinz~\cite{quasi2}.
These analyses indicate that the lattice results can be fit surprisingly
well by an ideal gas of massive quarks and 
gluons with temperature-dependent masses 
$m_q(T)$ and $m_g(T)$ that grow approximately linearly with $T$.

A straightforward application of screened perturbation 
theory to gauge theories like QCD would be doomed to failure, 
because a local mass term for gluons 
is not gauge invariant.  
There is a way to incorporate plasma effects, including quasiparticle masses,
screening of the gauge interaction and Landau damping, into perturbative calculations
while maintaining gauge invariance and that is by using hard-thermal-loop
(HTL) perturbation theory.
Hard-thermal-loop perturbation theory was originally developed to sum
up all higher loop corrections that are leading order in $g$ for amplitudes
having soft external lines with momenta of order $gT$~\cite{Braaten-Pisarski}. 
If it is applied
to amplitudes with hard external lines with momenta of order $T$, it 
selectively resums corrections that are higher order in $g$.
This resummation 
is a generalization of screened
perturbation theory that respects gauge invariance.
It corresponds essentially to expanding around an
ideal gas of quark and gluon quasiparticles. 

HTL perturbation theory is defined by
adding and
subtracting hard-thermal-loop correction terms to the action
\cite{Braaten-Pisarski}. The HTL correction terms are nonlocal, 
and the resulting effective
propagators and vertices are complicated functions of the energies and
momenta.  
Calculations in HTL perturbation theory are therefore much more difficult
than in screened perturbation theory.
The nonlocality of
the HTL correction terms also introduces conceptual problems
associated with renormalization.  Since
these terms are not renormalizable in the standard sense, the general structure
of the ultraviolet divergences that they generate is not known.

If HTL perturbation theory proves to be tractable and if issues associated
with renormalization can be resolved, it could have very important 
applications.
It is 
a reorganization of QCD perturbation
theory around a starting point that is essentially an ideal gas of
massive quasiparticles. In contrast to quasiparticle models,
the corrections
due to interactions between quasiparticles
can be calculated systematically, at least through
next-to-next-to-leading order in HTL perturbation theory.
Beyond 
that order perturbative calculations must be supplemented by a
nonperturbative method to deal with the magnetic mass problem.
A significant advantage of HTL perturbation theory over the approach
using lattice gauge theory
calculations in DRQCD is that it can be readily applied to the real-time
processes that are the most promising signatures of a quark-gluon plasma.

In this paper, we calculate the free energy of a hot gluon plasma explicitly
to leading order in hard-thermal-loop perturbation theory.  In spite
of the complexity of the HTL propagators, their analytic properties can be 
used to make explicit calculations 
possible.  Although complicated ultraviolet divergences arise in the
calculation, the most severe divergences cancel between quasiparticle and 
Landau-damping
contributions and between transverse and longitudinal contributions.  The
remaining divergences arise from integrating over large three-momentum 
and can be
removed by a counterterm proportional to $m_g^4$
at the expense of introducing a renormalization scale.  
With reasonable choices of the renormalization scales, 
our leading order result for the HTL free energy lies 
below results from lattice QCD for $T>2T_c$.
However, the deviation from lattice QCD results
has the correct sign and roughly the correct magnitude to be
accounted for by next-to-leading order corrections in HTL perturbation theory.

\section{HTL Resummation}
We begin this section 
by defining screened perturbation theory for a
scalar field theory. We then define hard-thermal-loop perturbation theory, 
which is a
generalization of screened perturbation theory 
that respects gauge invariance.
Finally, we write down a formal expression for the free energy at 
leading order in HTL
perturbation theory.
\subsection{Screened Perturbation Theory}
The lagrangian density for a massless scalar field with a $\phi^4$
interaction is
\bqa
{\cal L}={1\over2}\partial_{\mu}\phi\partial^{\mu}\phi
-{1\over 24}g^2\phi^4+\Delta{\cal L}(g^2),
\eqa
where $g$ is the coupling constant and $\Delta{\cal L}$ includes counterterms.
The conventional perturbative expansion in powers of $g^2$ generates
ultraviolet divergences, and the counterterm $\Delta{\cal L}$
must be adjusted to cancel the divergences order by order in $g^2$.
At nonzero temperature, the conventional perturbative expansion also generates
infrared divergences. They can be removed by resumming the higher order 
diagrams that generate a thermal mass of order $gT$ for the scalar particle.
This resummation changes the perturbative series from an expansion in powers
of $g^2$ to an expansion in powers of $\left(g^2\right)^{1/2}=g$.
Unfortunately, the perturbative expansion for the free energy has 
large coefficients and
appears to be convergent only for tiny values of $g$\cite{Parwani}.

Screened perturbation theory, which was introduced  by Karsch, Patk\'os
and Petreczky~\cite{K-P-P}, is simply a reorganization of the perturbation
series for thermal field theory.
It can be made more systematic by using a framework called 
``optimized perturbation theory'' that Chiku and Hatsuda~\cite{Chiku-Hatsuda}
have applied to a spontaneously broken scalar field theory. The Lagrangian
density is written as
\bqa
\label{SPT}
{\cal L}={\cal L}-{1\over2}(m^2-\delta m^2)\phi^2
+\Delta{\cal L}_{\rm s}(g^2,m^2-\delta m^2),
\eqa
where $\delta=1$ will be used as a formal expansion parameter and 
$\Delta{\cal L}_{\rm s}$ includes the additional counterterms that are required
to remove ultraviolet divergences for $\delta\neq 1$.
If we simply set $\delta=1$, the additional terms in~(\ref{SPT}) vanish.
Screened perturbation theory is defined by taking $m^2$ to be of order 
$g^0$ and $\delta$ to be of order $g^2$, expanding systematically in powers
of $g$ and setting $\delta=1$ at the end of the calculation.
This defines a reorganization of perturbation theory in which the expansion
is around the free field theory defined by
\bqa
\label{freesca}
{\cal L}_{\rm free}={1\over2}\partial_{\mu}\phi\partial^{\mu}\phi
-{1\over2}m^2\phi^2.
\eqa
The effects of the $m^2$ term in~(\ref{freesca}) are included to all 
orders, but they are systematically subtracted out at higher
orders in perturbation theory by the $\delta m^2$ term in~(\ref{SPT}).

This reorganization of perturbation theory generates new ultraviolet
divergences, but they can be canceled by the additional counterterms
in $\Delta{\cal L}_{\rm s}$. The renormalizability of the lagrangian 
in~(\ref{SPT}) guarantees that the only counterterms that are required
are $\partial_{\mu}\phi\partial^{\mu}\phi$, $\phi^2$, and $\phi^4$.
At nonzero temperature, screened perturbation theory does not generate
any infrared divergences, because the mass parameter $m^2$ in the
free lagrangian~(\ref{freesca}) provides an infrared cutoff. 
The resulting perturbative expansion is therefore a power series in $g^2$
whose coefficients depend on the mass parameter $m^2$.

The parameter $m^2$ in screened perturbation theory is completely arbitrary.
To complete a calculation in screened perturbation theory, it is necessary
to specify $m^2$ as a function of $g^2$ and $T$. 
Karsch, Patk\'os, and Petreczky used the solution of a gap equation
as their prescription for $m^2(T)$. The resulting
loop expansion for the free energy appeared to be convergent.

In the weak coupling limit $g\to 0$, the solution to the gap 
equation for $m^2(T)$ approaches $g^2T^2/24$. If we use this
value for $m^2$ and then reexpand the perturbative series for the free energy
in powers of $g$, we recover the conventional perturbative expansion in 
powers of $g$, with its lack of convergence.
It is therefore essential to keep the full dependence of $m^2$
at every order in $g^2$ and not expand in powers of $m/T$.

\subsection{HTL Perturbation Theory}
Renormalized perturbation theory for pure-glue QCD can be defined by expressing
the QCD lagrangian density in the form
\bqa
{\cal L}_{\rm QCD}=-{1\over2}{\rm Tr}\left(G_{\mu\nu}G^{\mu\nu}\right)
+{\cal L}_{\rm gf}+{\cal L}_{\rm ghost}+\Delta{\cal L}_{\rm QCD},
\eqa
where $G_{\mu\nu}=\partial_{\mu}A_{\nu}-\partial_{\nu}A_{\mu} -ig[A_{\mu},A_{\nu}]$ is the gluon field strength and $A_{\mu}$ is the gluon field expressed
as a matrix in the $SU(3)$ algebra.
The ghost term ${\cal L}_{\rm ghost}$ depends on the choice of 
the gauge-fixing term ${\cal L}_{\rm gf}$. The perturbative expansion in powers
of $g$ generates ultraviolet divergences, and the counterterms in 
$\Delta{\cal L}_{\rm QCD}$ are adjusted to cancel those divergences order by order
in $g$. 

Hard-thermal-loop (HTL) perturbation theory is simply a reorganization of the perturbation 
series for thermal QCD. The lagrangian density is written as 
\bqa
\label{QCDL}
{\cal L}_{\rm QCD}={\cal L}_{\rm QCD}+{\cal L}_{\rm HTL}+\Delta{\cal L}_{\rm HTL}(g,m_g^2-\delta m_g^2).
\eqa
The HTL improvement term is 
\bqa
\label{impr}
{\cal L}_{\rm HTL}=-{3\over2}(m_g^2-\delta m_g^2){\rm Tr}
\left(G_{\mu\alpha}\left\langle {n^{\alpha}n^{\beta}\over(n\cdot D)^2}\right\rangle_{\!\!n}G^{\mu}_{\;\;\beta}\right),
\eqa
where 
$D_{\mu}$ is the covariant derivative in the adjoint representation,
$n^{\mu}=(1,\hat{{\bf n}})$
is a light-like four-vector, $\langle\ldots\rangle_{ n}$ 
represents the average over the directions
of $\hat{{\bf n}}$, and $\delta=1$ is a formal expansion parameter.
The term~(\ref{impr}) is the effective lagrangian that would be induced by
a rotationally invariant ensemble of colored sources with infinitely high
momentum. The parameter $m_g$ can be identified with the plasma frequency,
or equivalently, with the rest mass of a gluon quasiparticle.

If we set $\delta=1$, the coefficient of the HTL improvement term~(\ref{impr})
vanishes. 
HTL perturbation theory is defined by 
taking $m_g^2$ to be of order $g^0$ and $\delta$ to be formally of 
order $g^2$, expanding systematically in powers of $g$, and then setting 
$\delta=1$ at the end of the calculation.
This defines a reorganization of the perturbative series
in which some of the effects of  
the $m_g^2$ term in~(\ref{impr})
are included to all orders but then systematically subtracted out
at higher orders in perturbation theory by the
$\delta m^2_g$ term. If the perturbative corrections from the
$\delta m_g^2$ 
term could be summed to all orders, there would be no dependence
on $m_g$. However, any truncation of the expansion in $\delta$ produces results
that depend on $m_g$.

The term $\Delta{\cal L}_{\rm HTL}$ in~(\ref{impr}) includes
the counterterms required to cancel the new ultraviolet divergences generated
by the reorganization of the perturbative series. 
Unlike the counterterms $\Delta{\cal L}_{\rm QCD}$ in~(\ref{QCDL}) which are
constrained by locality and renormalizability, the general structure
of the terms in $\Delta{\cal L}_{\rm HTL}$ is unknown.
We anticipate that if we use a scale-invariant regularization method
such as dimensional regularization, the terms in $\Delta{\cal L}_{\rm HTL}$
will be polynomials in the parameters $g$ and $m_g^2-\delta m_g^2$.

The free lagrangian that serves as a starting point for HTL perturbation theory
is obtained by setting $g=0$ and $\delta=0$ in~(\ref{QCDL}).
Choosing a covariant gauge-fixing term with gauge parameter $\xi$, 
the free lagrangian is
\bqa\nonumber
{\cal L}_{\rm free}&=&-{\rm Tr}\left(\partial_{\mu}A_{\nu}
\partial^{\mu}A^{\nu}-\partial_{\mu}A_{\nu}\partial^{\nu}A^{\mu}\right)
-{1\over\xi}{\rm Tr}\left(\left(\partial^{\mu}A_{\mu}\right)^2\right) \\
\label{covfree}
&&-{3\over2}m_g^2\mbox{Tr}
\left((\partial_{\mu}A_{\alpha}-\partial_{\alpha}A_{\mu})
\left\langle{n^{\alpha} n^{\beta}\over(n\cdot\partial)^2}\right\rangle_{\!\!n}
(\partial^{\mu}A_{\beta}-\partial_{\beta}A^{\mu})\right).
\eqa
The HTL improvement term generates a self-energy tensor
for the gluon that is diagonal in the color indices and has the form
\bqa
\label{glue}
\Pi^{\mu\nu}(K)=-{3\over2}m_g^2\left\langle
{K^2n^{\mu}n^{\nu}-n\cdot K(K^{\mu}n^{\nu}+n^{\mu}K^{\nu})+(n\cdot K)^2g^{\mu\nu}\over(K\cdot n)^2}
\right\rangle_{\!\!n},
\eqa
where $K^{\mu}=(k_0,{\bf k})$ is the Minkowski four-momentum and
$K^2=k_0^2-{\bf k}^2$. The HTL self-energy tensor
satisfies the Ward identity $K_{\mu}\Pi^{\mu\nu}=0$.
Because of this Ward identity and the rotational symmetry
around the axis $\hat{{\bf k}}$, the tensor $\Pi^{\mu\nu}$
can be expressed in terms of two independent functions 
$\Pi_L$ and $\Pi_T$:
\bqa
\Pi^{\mu\nu}(K)
=-\Pi_L(K){K^2g^{\mu\nu}-K^{\mu}K^{\nu}\over k^2}
+\left[\Pi_T(K)-{K^2\over k^2}\Pi_L(K)\right]
g^{\mu i}(\delta^{ij}-\hat{k}^i\hat{k}^j)g^{j\nu}.
\eqa
The longitudinal and transverse self-energies are
\bqa
\Pi_L(K)&=&\Pi^{00}(K),\\ 
\Pi_T(K)&=&={1\over2}\left(\delta^{ij}-\hat{k}^i\hat{k}^j\right)\Pi^{ij}(K).
\eqa
The inverse propagator for the gluon is diagonal in the 
color indices and has the form
\bqa\label{gpr}
\Delta^{-1}(K)^{\mu\nu}&=&-K^2g^{\mu\nu}+\left(1-{1\over\xi}\right)K^{\mu}K^{\nu}
+\Pi^{\mu\nu}(K).
\eqa
If there are $d$ spatial dimensions, the
matrix $\Delta^{-1}(K)^{\mu}_{\;\;\nu}$
has $d+1$ eigenvalues: $-K^2(k^2-\Pi_L)/k^2$, $-K^2/\xi$,
and the $(d-1)$-fold degenerate eigenvalue $K^2-\Pi_T$.
The inverse propagator for the ghost in a general covariant gauge is
$\Delta^{-1}_{\rm ghost}=K^2$.

In the general Coulomb gauge, the gauge-fixing term in~(\ref{covfree})
is replaced by
$-(1/\xi){\rm Tr}\left(\left(\nabla\cdot {\bf A}\right)^2\right)$.
The term $-(1/\xi)K^{\mu}K^{\nu}$ in the inverse gluon propagator~(\ref{gpr})
is replaced by $-(1/\xi)g^{\mu i}k^ik^jg^{j\nu}$.
The $d+1$ eigenvalues of $\Delta^{-1}(K)^{\mu}_{\;\;\nu}$
are the $(d-1)$-fold degenerate eigenvalue
$K^2-\Pi_T$ and two other eigenvalues whose
product is $-k^2(k^2-\Pi_L)/\xi$. The inverse propagator for the ghost
in this gauge is $k^2$.
The limit $\xi\rightarrow 0$
is the strict Coulomb gauge defined by the constraint $\nabla\cdot{\bf A}=0$.
In this gauge, the only propagating modes of the gluon are $A_0$
and the $d-1$ transverse components of ${\bf A}$.
The propagators are $k^2-\Pi_L$ for $A_0$ and
$K^2-\Pi_T$ for the transverse components of ${\bf A}$.
\subsection{HTL Free Energy}
In the imaginary-time formalism, the renormalized
one-loop free energy can be written as
\bqa
\label{trlog}
{\cal F}_{\rm HTL}=(N_c^2-1)\left[{1\over2}\sumint_K{\rm Tr}\log\Delta^{-1}(K)
-\sumint_K\log\Delta_{\rm ghost}^{-1}(K)
+\Delta{\cal F}\right],
\eqa
where $\Delta^{-1}$ and ${\Delta}^{-1}_{\rm ghost}$ are the 
euclidean inverse propagators for gluons and ghosts, and
$K$ is now a euclidean four-momentum: $K^{\mu}=(\omega_n,{\bf k})$, 
and $K^2=k^2+\omega^2_n$.
The sum-integral in~({\ref{trlog}) represents a dimensionally regularized integral over the
momentum ${\bf k}$ and a sum over the Matsubara frequencies $\omega_n=2\pi nT$:
\bqa
\sumint_K\equiv T \sum_{n=-\infty}^{\infty}\mu^{3-d}\int{d^d k\over(2 \pi)^d}.
\eqa
The factor of $\mu^{3-d}$, where $\mu$ is a renormalization scale,
ensures that 
the regularized free energy has 
correct dimensions even for $d\neq 3$. The counterterm $\Delta{\cal F}$
in~(\ref{trlog}) can be used to cancel ultraviolet divergences that have
the form of an additive constant in the free energy.

In the gluon term in~(\ref{trlog}), the integrand is the sum of the
logarithms of the $d+1$ eigenvalues of $\Delta^{-1}$. 
After taking into account the cancellations from 
the ghost term and dropping terms that vanish in dimensional regularization,
the expression~(\ref{trlog}) for the HTL free energy reduces to
\begin{equation}
\label{freedef}
{\cal F}_{\rm HTL} \;=\; (N_c^2-1)
\left[(d-1) {\cal F}_T + {\cal F}_L 
	\;+\; \Delta {\cal F} \right],
\end{equation}
where
\begin{eqnarray}  
{\cal F}_T & = & {1 \over 2}\sumint_K \log [K^2 + \Pi_T(K)],
\label{FT-def}\\
{\cal F}_L & = & {1 \over 2}\sumint_K \log [k^2 - \Pi_L(K)].
\label{FL-def}
\end{eqnarray}
An identical expression for the free energy is obtained in the
general Coulomb gauge.
The free energies ${\cal F}_T$ and ${\cal F}_L$ have simple interpretations in the strict Coulomb gauge.
They are simply the contributions to the free energy
from a transverse component
of ${\bf A}$ and from $A_0$, respectively.
We therefore refer to them as the transverse and longitudinal free 
energies.

At first sight, the calculation of the HTL free energy appears to be a rather
daunting mathematical problem.
The HTL self-energy functions in
(\ref{FT-def}) and (\ref{FL-def}) are
\begin{eqnarray}
\Pi_T(\omega_n,k) & = & - {3 \over 2} m_g^2 {\omega_n^2 \over k^2} 
\left[1 + {\omega^2_n + k^2 \over 2i\omega_nk} 
	\log{i\omega_n+k \over i\omega_n - k} \right],
\label{Pi-T}\\
\Pi_L (\omega_n,k) &=& 3m_g^2 \left[{i \omega_n \over 2k} 
	\log {i\omega_n +k \over i \omega_n-k} - 1 \right].
\label{Pi-L}
\end{eqnarray}
In order to calculate the HTL free energy, we must evaluate the sum-integrals
in~(\ref{FT-def}) and~(\ref{FL-def}) with a regularization that cuts off
the ultraviolet divergences.
In the next section, we show that 
this problem can be made tractable by using the analytic properties of 
the HTL self-energies.
\section{HTL Free Energy}
In this section, we reduce the HTL free energy to integrals that can be 
evaluated numerically. We first compute the free energy of a free massless
boson as a simple example.
We then compute the transverse and longitudinal free energies, separating them
into quasiparticle and Landau-damping terms and isolating the ultraviolet
divergences into integrals that can be computed analytically. The most
severe divergences cancel, leaving a logarithmic divergence that must be
canceled explicitly by a counterterm.
\subsection{Free Massless Boson}
As a warm-up exercise for computing the transverse free energy, we compute
the free energy of a free massless boson:
\bqa
\label{freebos}
{\cal F}_{\rm boson}={1\over2}\sumint_K\log(K^2).
\eqa
In dimensional regularization, the $n=0$ term in the sum-integral vanishes.
For the same reason, we can subtract $\log k^2$ from the integrand for
each of the $n\neq 0$ terms. The free energy becomes
\bqa
\label{freebos2}
{\cal F}_{\rm boson}={1\over2}T\sum_{n\neq 0}\int_{\bf k}
\log{k^2+\omega_n^2\over k^2}.
\eqa
We have introduced a compact notation for the dimensionally regularized
integral over momentum together with the renormalization scale factor:
\bqa 
\int_{\bf k}\equiv \mu^{3-d}\int{d^dk\over(2\pi)^d}.
\eqa
The sum over Matsubara frequencies in~(\ref{freebos2}) can be 
expressed as a contour integral:
\bqa
\label{freecon}
{\cal F}_{\rm boson}={1\over2}\int_{C}{d\omega\over2\pi i}
\int_{\bf k}\log{k^2-\omega^2\over k^2}{1\over e^{\beta\omega}-1},
\eqa 
where the contour $C$ encloses the points $\omega=i\omega_n$, $n\neq 0$,
as shown in Fig.~\ref{boson}.

\begin{figure}[htb]


\hspace{1cm}
\epsfysize=8cm
\centerline{\epsffile{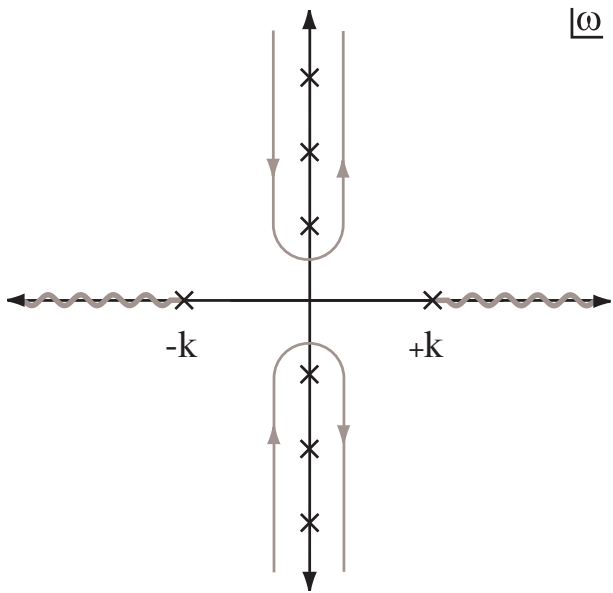 }}
\\
\caption[a]{The free energy of a massless boson can be expressed as an
integral over a contour $C$ that can be deformed into two contours 
that wrap around
the branch cuts of $\log(k^2-\omega^2)$.}
\label{boson}
\end{figure}
We can choose the branch cuts of the logarithm to run along the real 
$\omega$-axis from $-\infty$ to $-k$ and from $+k$ to $+\infty$.
It is convenient to insert a convergence factor 
$e^{\eta\omega}$ ($\eta\rightarrow 0^+$) into the integrand. This allows the contour $C$ in Fig.~\ref{boson} to be deformed into two contours that wrap around the 
branch cuts. 
Collapsing the contour onto the two branch cuts, 
making the change of variables $\omega\to-\omega$ in the contribution from 
the negative branch cut,
and taking
the limit $\eta\rightarrow 0$ whenever it is not needed 
for convergence, the free energy reduces to
\bqa
{\cal F}_{\rm boson}=-\int_{\bf k}\int_k^{\infty}d\omega
\left(
{1\over e^{\beta\omega}-1}+{1\over2}e^{-\eta\omega}\right).
\eqa
Integrating over $\omega$ and taking the limit $\eta\rightarrow 0$, we obtain
\bqa
\label{bosfin}
{\cal F}_{\rm boson}=\int_{\bf k}
\left[T\log(1-e^{-\beta k}) -{1\over2\eta}+{1\over2}k\right].
\eqa
The integral over the last two terms are zero in dimensional regularization.
Setting $d=3$ in the integral of the first term in~(\ref{bosfin}) 
and evaluating it analytically, we
get 
\bqa
{\cal F}_{\rm boson}=-{\pi^2\over90}T^4. 
\eqa
Summing over the $N_c^2-1$
color degrees of freedom and the two spin degrees of freedom of a transverse
gluon, we obtain the free energy of an ideal gas of massless gluons:
\bqa
{\cal F}_{\rm ideal}=-\left(N_c^2-1\right){\pi^2\over 45}T^4.
\eqa
\subsection{Transverse Free Energy}
We now proceed to compute the transverse free energy in~(\ref{FT-def}). 
Following the
steps that led from (\ref{freebos}) to (\ref{freecon}), we obtain the
contour integral
\bqa
\label{transcont}
{\cal F}_{T}={1\over2}\int_C{d\omega\over2\pi i}\int_{\bf k}
\log{k^2-\omega^2+\Pi_T(-i\omega,k)\over k^2}{e^{\eta\omega}
\over e^{\beta\omega}-1},
\eqa
where $e^{\eta\omega}$ ($\eta\rightarrow 0^+$) is a convergence factor
and the contour $C$ encloses the points $\omega=i\omega_n,n\neq 0$.
The integrand has logarithmic branch cuts in $\omega$ that run along the real
$\omega$-axis from 
$-\infty$ to $-\omega_T(k)$
and from $+\omega_T(k)$ to $+\infty$,
where $\omega_T(k)$ is the quasiparticle
dispersion relation for transverse gluons. This dispersion relation satisfies
$k^2-\omega_T^2+\Pi_T(-i\omega_T,k)=0$, or
\bqa
\omega_T^2 & = & k^2 \;+\; {3 \over 2} m_g^2 {\omega_T^2 \over k^2} 
\left[ 1 - {\omega_T^2 - k^2 \over 2 \omega_T k} 
		\log {\omega_T + k \over \omega_T - k} \right].
\label{omega-T}
\eqa
The integrand also has a branch cut in $\omega$ running from $-k$ to $+k$
due to the function $\Pi_T$. This branch cut represents the effects of
Landau damping. 
The contour $C$ can be deformed into a contour that wraps around the three
branch cuts as in Fig.~\ref{trans1}.

\begin{figure}[htb]


\hspace{1cm}
\epsfysize=8cm
\centerline{\epsffile{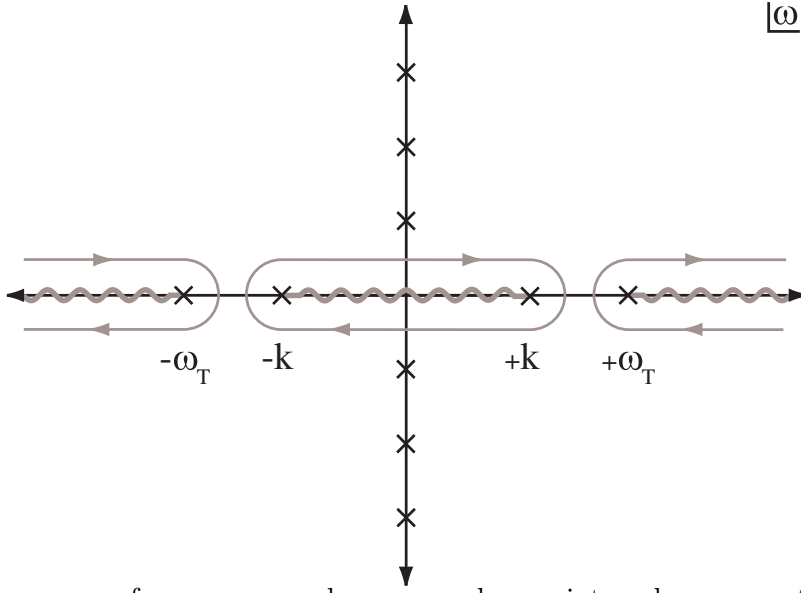}}
\caption[a]{The transverse free energy can be expressed as an integral over
a contour $C$ that wraps around three branch cuts
of $\log(k^2-\omega^2+\Pi_T)$.}
\label{trans1}
\end{figure}
We identify the contribution from the 
branch cuts ending at $\pm\omega_T(k)$ as the quasiparticle part of 
${\cal F}_T$ and denote it by ${\cal F}_{T,{\rm qp}}$. Following the same 
steps that led to~(\ref{bosfin}), we obtain
\bqa
\label{T-qp}
{\cal F}_{T,{\rm qp}}=
\int_{\bf k}
\left[T\log(1-e^{-\beta \omega_T})+{1\over2}\omega_T\right].
\eqa
We have omitted the term $1/(2\eta)$ in the integrand of~(\ref{T-qp})
because its integral
vanishes in dimensional regularization.
We identify the contribution from the contour wrapping around the branch cut
running from
$-k$ to $+k$ as the Landau-damping part of ${\cal F}_T$ and denote it
by ${\cal F}_{T,{\rm Ld}}$.
Collapsing the contour onto the branch cut and expressing it as an
integral over positive values of $\omega$, we obtain
\bqa
\label{T-Ld2}
{\cal F}_{T,{\rm Ld}}=
-{1\over\pi}\int_{\bf k}\int_{0}^kd\omega\;
\phi_T\left[{1\over e^{\beta\omega}-1}+{1\over2}\right].
\eqa
The angle $\phi_T$ is 
\bqa
\phi_T(\omega,k)=
\arctan{{3 \pi \over 4} m^2_g {\omega K^2 \over k^3}\over 
K^2 \;+\; {3 \over 2} m^2_g {\omega^2 \over k^2}
\left[ 1 + {K^2 \over 2 k \omega} L \right]},
\label{theta-T} 
\eqa
where $K^2=k^2-\omega^2$ and 
$L=\log\left[{(k+\omega)/(k-\omega)}\right]$.
The angle $\phi_T(\omega,k)$ vanishes as $\omega\rightarrow 0$ and as 
$k\rightarrow \infty$ and it also vanishes at
$\omega=k$. 
The complete transverse energy ${\cal F}_T$ is the sum of the quasiparticle
term (\ref{T-qp}) and the Landau-damping term (\ref{T-Ld2}).
The integral of the second term in~(\ref{T-qp}) and the integral of both
terms in~(\ref{T-Ld2}) are ultraviolet divergent, but the divergences 
are regularized by the $d$-dimensional integral over ${\bf k}$.
In order to extract the divergences  analytically, we make subtractions
in the integrands that render the integrals 
finite in $d=3$ dimensions and then 
extract the poles in $d-3$ from the subtracted integrals.

We first consider the transverse quasiparticle term~(\ref{T-qp}).
The integral of the second term is divergent, because
the asymptotic form of the transverse dispersion relation
for large $k$ is~\cite{pisarski1}
\bqa
\omega_T(k)\longrightarrow k+{3\over4}{m_g^2\over k}
-{9\over32}{m_g^4\over k^3}\left(2\log{8k^2\over3m_g^2}-3\right).
\eqa
Our subtraction should make the integral ultraviolet convergent for $d=3$,
and it should
not introduce any infrared divergences. Our choice for the subtracted integral
is
\bqa
\label{T-qp-s}
{\cal F}_{T,{\rm qp}}^{(\rm sub)}=
{1 \over2}\int_{\bf k}\left[ 
\sqrt {k^2 + \mbox{${3 \over 2}$} m_g^2}
- {9 m_g^4 \over 16(k^2 + {3 \over 2} m^2_g )^{3/2}}
\left( \log {8(k^2 + {3 \over 2} m_g^2) \over 3 m_g^2} -2 \right) \right ].
\eqa
After subtracting this from~(\ref{T-qp}), we can take the limit 
$d\rightarrow 3$:
\bqa
{\cal F}_{T,{\rm qp}}-{\cal F}_{T,{\rm qp}}^{(\rm sub)} & = & 
{T \over 2 \pi^2} \int^\infty_0 k^2 dk \log (1-e^{-\beta \omega_T})
\nonumber 
\\ 
&\hspace{-1cm}&\hspace{-2cm} +{1 \over 4 \pi^2} \int^\infty_0 k^2 dk \left[ \omega_T 
	- \sqrt {k^2 + \mbox{${3 \over 2}$} m_g^2}
+ {9 m_g^4 \over 16(k^2 + {3 \over 2} m^2_g )^{3/2}}
\left( \log {8(k^2 + {3 \over 2} m_g^2) \over 3 m_g^2} -2 \right) \right ].
\label {FT-qp}
\eqa 
If we impose a momentum cutoff $k<\Lambda$, our subtraction 
integral~(\ref{T-qp-s}) has power divergences proportional to $\Lambda^4$
and $m_g^2\Lambda^2$ and logarithmic divergences proportional to 
$m_g^4\log\Lambda$ and $m_g^4\log^2\Lambda$.
The quartic divergence is canceled by the usual renormalization of the
vacuum energy density at zero temperature.
Dimensional regularization throws away the power divergences and replaces
the logarithmic divergences by poles in $d-3$. 
In the limit $d\rightarrow 3$, the individual integrals
in~(\ref{T-qp-s}) are given by~(\ref{qpsub1})--(\ref{qpsub3}) in the appendix.
The result is
\bqa
\label{T-qp-f}
{\cal F}_{T,\rm qp}^{\rm (sub)}=
-{9\over64}m_g^4\left(3m_g^2\over2\mu^2\right)^{-\epsilon}
{\Omega_d\over(2\pi)^d}
\left[{1\over\epsilon^2}+{4\log2-3\over2}{1\over\epsilon}
+2\log^22-3\log2+{\pi^2\over6}-{1\over4}\right],
\eqa
where $\epsilon=(3-d)/2$ and $\Omega_d=2\pi^{d/2}/\Gamma(d/2)$
is the angular integral in $d$ dimensions.

We next consider the Landau-damping term~(\ref{T-Ld2}). Again we must
choose a subtraction term that removes the ultraviolet divergences without
introducing infrared divergences. Our choice is
\bqa\nonumber
{\cal F}_{T,\rm Ld}^{(\rm sub)}&=&-{1\over\pi}\int_{\bf k}
\int_0^{k}d\omega\left[{3\pi m_g^2\omega\over4k^3}{1\over e^{\beta\omega}-1}+
{3\pi m_g^2\omega K^2\over8k^3(K^2+{3\over2}m_g^2)}
\right.\\
\label{T-Ld-s}
&&\left.
\hspace{3.5cm}-{9\pi m_g^4\omega (K^2)^2\over16k^5(K^2+{3\over2}m_g^2)^2}
\left({\omega\over2k}L-1\right)\right].
\eqa
After subtracting this from~(\ref{T-Ld2}), we can take the limit 
$d\rightarrow 3$:
\bqa\nonumber
{\cal F}_{T,{\rm Ld}}-{\cal F}_{T,{\rm Ld}}^{(\rm sub)} & = & 
- {1 \over 2 \pi^3} \int^\infty_0 d\omega
{1 \over e^{\beta \omega} -1} \int^\infty_\omega k^2 dk 
\left [ \phi_T - {3 \pi m_g^2\omega \over 4 k^3} \right]\\ 
&&\hspace{-2.03cm}-{1 \over 4 \pi^3} \int^\infty_0 d\omega 
\int^\infty_\omega k^2 dk 
\Bigg[ \phi_T - {3 \pi m_g^2 \omega K^2 \over 4 k^3
(K^2 + {3 \over 2} m_g^2)}
 + {9 \pi m_g^4 \omega (K^2)^2 \over 8 k^5 (K^2 + {3 \over 2} m_g^2)^2}
\left ( { \omega \over 2k} L - 1 \right)
\Bigg].
\label{FT-Ld} 
\eqa
In both integrals, the first subtraction 
makes them convergent as $k\rightarrow\infty$
with $\omega$ fixed. The second subtraction in the second integral
makes it convergent as $k\rightarrow \infty$ with $\omega\sim k$.
If we impose ultraviolet
cutoffs $\omega<\Lambda$ and $k<\Lambda$ on the energy and
momentum, the subtraction integral~(\ref{T-Ld-s}) has a power divergence
proportional to $m_g^2\Lambda^2$ and logarithmic
divergences proportional to $m_g^2T^2\log\Lambda$, $m_g^4\log\Lambda$,
and $m_g^4\log^2\Lambda$.
The divergence proportional to
$m_g^4\log^2\Lambda$
cancels against the corresponding divergence
in the quasiparticle subtraction
integral~(\ref{T-qp-s}). 
The cancellation can be traced to the fact that $\Pi_T(\omega,k)$
is analytic in the variable $\omega$ at $\omega=\infty$.
Dimensional regularization replaces the logarithmic divergences 
in~(\ref{T-Ld-s}) by poles in $d-3$ and sets the power divergences
to zero. The integrals in~(\ref{T-Ld-s}) 
are evaluated in the limit $d\rightarrow 3$ in the appendix 
and given by~(\ref{Ldtr1})--(\ref{Ldtr3}) and (\ref{Ldtemp1}). The result is
\bqa\nonumber
{\cal F}_{T,\rm Ld}^{(\rm sub)}
&=&\nonumber
-{\pi^2\over16}m_g^2T^2\left({T^2\over\mu^2}\right)^{-\epsilon}{\Omega_d\over(2\pi)^d}
\left[{1\over\epsilon}+2{\zeta^{\prime}(-1)\over\zeta (-1)}-2\log(2\pi)\right]\\
&&
\label{T-Ld-f}\hspace{-1.6cm}
+{9\over64}m_g^4\left({3m_g^2\over2\mu^2}\right)^{-\epsilon}{\Omega_d\over(2\pi)^d}\left[
{1\over\epsilon^2}+{2(\log2-1)\over3}{1\over\epsilon}+{2\over3}\log^22
-{22\over9}\log2+{\pi^2\over6}+{10\over9}
\right].
\eqa
Our final result for the transverse self-energy is the sum 
of~(\ref{FT-qp}),~(\ref{T-qp-f}),~(\ref{FT-Ld}) and~(\ref{T-Ld-f}).
Note that the divergence proportional to ${1/\epsilon^2}$
cancels between~(\ref{T-qp-f}) and~(\ref{T-Ld-f}). 

\subsection{Longitudinal Free Energy}
We now compute the longitudinal free energy given 
in~(\ref{FL-def}). 
We first separate out the $n=0$ term:
\bqa
\label{sepsum}
{\cal F}_L={1\over2}T\int_{\bf k}\log\left({k^2+3m^2_g}\right)+
{1\over2}T\sum_{n\neq 0}\int_{\bf k}
\log\left({k^2-\Pi_L(\omega_n,k)}\right).
\eqa
These two terms can be expressed as a single integral 
over a contour $C$ that encloses the points $\omega=i\omega_n,n\neq 0$:
\bqa
\label{locont}
{\cal F}_L={1\over2}\int_C{d\omega\over2\pi i}\int_{\bf k}
\log{k^2-\Pi_L(-i\omega,k)\over k^2+3m_g^2}
{e^{\eta\omega}\over e^{\beta\omega}-1},
\eqa
where $e^{\eta\omega}$ ($\eta\rightarrow 0^+$) is a convergence factor.
The logarithm in the integrand
can be expressed as the difference of two logarithms.
The integral of the first logarithm reproduces the sum of the 
$n\neq 0$ terms in~(\ref{sepsum}). In the contour integral of the 
$\log(k^2+3m_g^2)$ term, the contour can be deformed to wrap around the
pole at $\omega=0$ of the Bose-Einstein factor.
By the residue theorem, this reproduces the $n=0$ term in~(\ref{sepsum}).

The integrand in~(\ref{locont})
has a
logarithmic branch cut in $\omega$ that runs from $-\omega_L(k)$ to
$+\omega_L(k)$, where $\omega_L(k)$ is the quasiparticle dispersion relation
for longitudinal gluons. This dispersion relation satisfies 
$k^2-\Pi_L(-i\omega_L,k)=0$ or
\bqa
0 &=& k^2 \;+\; 3 m_g^2 
\left [ 1 - {\omega_L \over 2 k} 
	\log {\omega_L +k \over \omega_L-k} \right] .
\label{omega-L}
\eqa
The integral also has a branch cut in $\omega$ running from $-k$ to $+k$
due to the function $\Pi_L$. This branch cut represents the effects of Landau
damping. We choose both branch cuts to run along the real axis so that they
overlap in the region $-k<\omega<k$. 

The contour $C$ can be deformed 
into a contour that wraps around the branch cuts from as in Fig.~\ref{long1}, 
and the contour can then 
be collapsed onto the branch cut.
\begin{figure}[htb]


\hspace{1cm}
\epsfysize=8cm
\centerline{\epsffile{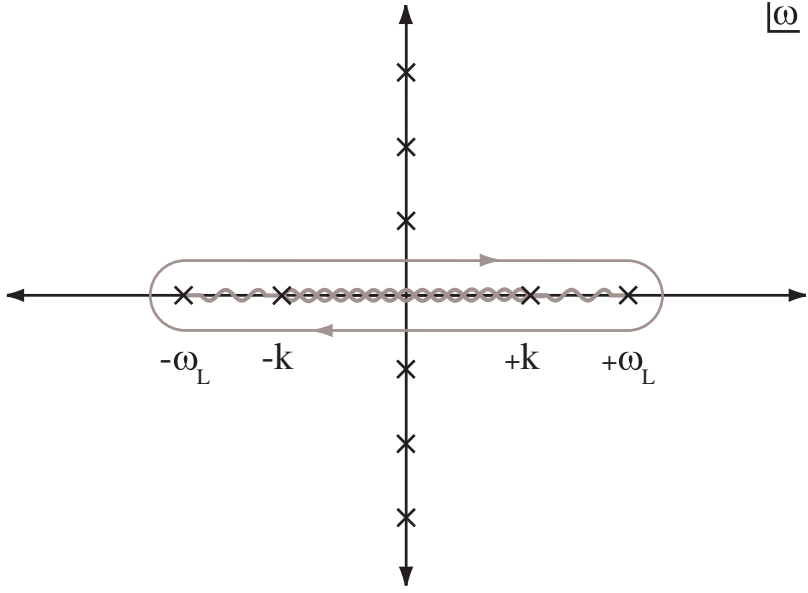}}
\\
\caption[a]{The longitudinal free energy can be expressed as an integral
around
a contour $C$ that wraps around the branch cuts of $\log(k^2-\Pi_L)$.}
\label{long1}
\end{figure}

We identify the contributions from $k<|\omega|<\omega_L(k)$ as the quasiparticle
part of ${\cal F}_L$ and denote it by ${\cal F}_{L,\rm qp}$:
\bqa
\label{L-qp}
{\cal F}_{L,\rm qp}=\int_{\bf k}
\left[T\log{1-e^{-\beta\omega_L}\over 1-e^{-\beta k}}
+{1\over2}\left(\omega_L-k\right)
\right].
\eqa
We identify the contribution from $|\omega|<k$ as the Landau-damping
part of ${\cal F}_L$ and denote it by ${\cal F}_{L,\rm Ld}$:
\bqa
\label{L-Ld}
{\cal F}_{L,\rm Ld}={1\over\pi}\int_{\bf k}\int_0^kd\omega\;
\phi_L\left[{1\over e^{\beta\omega}-1}+{1\over2}\right].
\eqa

The angle $\phi_L$ is 
\bqa
\phi_L(\omega,k)=\arctan{{3 \pi \over 2} m_g^2 {\omega \over k}\over
k^2 \;+\; 3m_g^2 
\left [ 1 - {\omega \over 2 k} L  \right]},
\label{theta-L}
\eqa
where $L=\log\left[(k+\omega)/(k-\omega)\right]$.
The angle $\phi_L(\omega,k)$ vanishes as $\omega\rightarrow0$ 
and as $k\rightarrow \infty$
and its value is $\pi$
at $\omega=k$. 
The denominator of the argument of the arctangent in~(\ref{theta-L})
vanishes at some point $\omega=\omega^*(k)$.
The arctangent in~(\ref{theta-L})
is the principal branch for $0<\omega<\omega^*(k)$,
and it jumps to the next branch
for $\omega^*(k)<\omega<k$,  
so that $\phi_L$ remains continuous.

The complete longitudinal free energy ${\cal F}_L$ is the 
sum of the quasiparticle term~(\ref{L-qp}) and the Landau-damping 
term~(\ref{L-Ld}). The quasiparticle integral is convergent, because
$\omega_L$ approaches $k$ as a gaussian at large $k$. The Landau-damping term
has ultraviolet divergences that are regularized by the $d$-dimensional
integral over $\bf k$. In order to extract the divergences analytically,
we make subtractions in the integrand that make the integral convergent in
$d=3$, and we then extract the poles in $d-3$ from the subtracted integral.
Our choice of the subtracted integral is
\bqa
\label{L-sub}
{\cal F}_{L,\rm Ld}^{(\rm sub)}&=&{1\over\pi}\int_{\bf k}
\int_0^kd\omega\left[
{3\pi m_g^2\omega\over2k^3}{1\over e^{\beta\omega}-1}+
{3\pi m_g^2\omega\over4k(k^2+3m_g^2)}
\right.
\left.+{9\pi m_g^4\omega^2\over8k^2(k^2+3m_g^2)^2}L\right].
\eqa
After subtracting this from~(\ref{L-Ld}), we can take the
limit $d\rightarrow 3$:
\bqa
{\cal F}_{L,{\rm Ld}}-{\cal F}_{L,\rm Ld}^{(\rm sub)}& = & 
{1 \over 2 \pi^3} \int^\infty_0 d\omega {1 \over e^{\beta \omega}-1} 
\int^\infty_\omega k^2 dk
\left [ \phi_L - { 3 \pi m_g^2 \omega \over 2k^3}
\right ]
\nonumber 
\\
&+& {1 \over 4 \pi^3} \int^\infty_0 d\omega 
\int^\infty_\omega  k^2 dk 
\left[ \phi_L 
- { 3 \pi m_g^2 \omega \over 2k (k^2 + 3m_g^2)}
- {9 \pi m_g^4 \omega^2 \over 4 k^2 (k^2 + 3m_g^2)^2} L \right].
\label {FL-Ld}
\eqa
If we impose ultraviolet cutoffs $\omega<\Lambda$ and $k<\Lambda$ on the
energy and momentum, the subtraction integral~(\ref{L-sub}) has a
power divergence proportional to $m^2_g\Lambda^2$ and logarithmic
divergences proportional to $m_g^2T^2\log\Lambda$ and $m_g^4\log\Lambda$.
The divergence proportional to $m_g^2\Lambda^2$ cancels against
the corresponding divergence in the transverse Landau-damping 
subtraction integral~(\ref{T-Ld-s}).
Dimensional regularization throws away the quadratic divergence and
replaces the logarithmic divergences by poles in $\epsilon$.
In the limit $d\rightarrow 3$, the individual integrals 
in~(\ref{L-sub}) are given by~(\ref{Ldlo1})--(\ref{Ldtemp1}) 
in the appendix. 
The result is
\bqa\nonumber
{\cal F}_{L,\rm Ld}^{(\rm sub)}
\nonumber
&=&{\pi^2\over8}m_g^2T^2\left({T^2\over\mu^2}
\right)^{-\epsilon}{\Omega_d\over(2\pi)^d}
\left[{1\over\epsilon}+2{\zeta^{\prime}(-1)\over\zeta(-1)}-2\log(2\pi)\right]\\
&&
\label{L-L-s}
+{3\over8}m_g^4\left({3m_g^2\over\mu^2}\right)^{-\epsilon}{\Omega_d\over(2\pi)^d}
\left[{\log2-1\over\epsilon}-\log2 -{1\over2}\right].
\eqa
Our final result for the longitudinal free energy is the sum 
of~(\ref{L-qp}),~(\ref{FL-Ld}) and~(\ref{L-L-s}). 
\subsection{Renormalized Free Energy}
To complete our calculation of the HTL free energy, we must determine the
counterterm $\Delta{\cal F}$ in~(\ref{freedef}) that cancels the 
divergences in $(d-1){\cal F}_T+{\cal F}_L$.
The ultraviolet divergences have been isolated in the subtraction terms~(\ref{T-qp-f})
and~(\ref{T-Ld-f}) for ${\cal F}_T$ and~(\ref{L-L-s}) for ${\cal F}_L$.
The poles in $\epsilon$ proportional to $m_g^2T^2$ cancel between
${\cal F}_T$ and ${\cal F}_L$. 
The cancellation can be traced to the fact that the logarithms 
cancels in the following combination of self-energy functions:
\bqa
2{\Pi_T(-i\omega,k)\over\omega^2-k^2}+{\Pi_L(-i\omega,k)\over k^2}
={3m_g^2\over\omega^2-k^2}.
\eqa
The remaining divergence 
is proportional to $m_g^4$. In the minimal
subtraction renormalization prescription, $\Delta{\cal F}$ is chosen to cancel
only the pole in $\epsilon$ and no additional finite terms:
\bqa
\label{df}
\Delta{\cal F}={9\over128\pi^2\epsilon}m_g^4.
\eqa
If we use a momentum cutoff $\Lambda$, there is also a quadratic divergence
proportional to $m_g^2\Lambda^2$
coming from the transverse quasiparticle subtraction term~(\ref{T-qp-s}). 
This would be canceled by an additional 
counterterm in~(\ref{df}) proportional to 
$m_g^2\Lambda^2$.

To obtain the final result for the HTL free energy, we insert 
into~(\ref{freedef}) the sum 
of~(\ref{FT-qp}),~(\ref{T-qp-f}),~(\ref{FT-Ld}) and~(\ref{T-Ld-f})
for ${\cal F}_T$, the sum of~(\ref{L-qp}),~(\ref{FL-Ld}) and~(\ref{L-L-s}) 
for
${\cal F}_L$, and~(\ref{df}) for $\Delta{\cal F}$. 
We can simplify the result by evaluating the temperature independent
integrals numerically. Since the only scale in these integrals is the
mass $m_g$, they reduce to $m_g^4$ multiplied by a numerical coefficient.
The contributions to ${\cal F}_T$ from the second integral in~(\ref{FT-qp}) 
and the second integral
in~(\ref{FT-Ld}) are $-8.656\times 10^{-3}m_g^4$ and 
$-4.554\times 10^{-4}m_g^4$, respectively.
The contribution to ${\cal F}_L$ from the integral of the
second term in~(\ref{L-qp})
and from the second integral in~(\ref{FL-Ld}) 
are $1.225\times 10^{-2}m_g^4$
and $4.290\times 10^{-3}m_g^4$, respectively.
Our final result for the renormalized free energy is
\bqa\nonumber
{\cal F}_{\rm HTL}&=&\left(N_c^2-1\right)\Bigg\{{1\over2\pi^2}T\int_0^{\infty}
k^2dk\left[2\log(1-e^{-\beta\omega_T})
+\log{1-e^{-\beta\omega_L}\over1-e^{-\beta k}}\right] \\ \nonumber
&&
\hspace{1.8cm}+{1\over2\pi^3}\int_0^{\infty}d\omega\;{1\over e^{\beta\omega}-1}
\int_{\omega}^{\infty}k^2dk\left[\phi_L-2\phi_T\right]\\
\label{total}
&&\hspace{1.8cm}+{1\over16}m_g^2T^2+{9\over64\pi^2}m_g^4
\left[\log{m_g\over\mu_3}-0.332837\right]\Bigg\},
\eqa 
where $\mu_3=\sqrt{4\pi}e^{-\gamma}\mu$ is the renormalization scale
associated with the modified minimal subtraction ($\overline{\mbox{MS}}$)
renormalization prescription and $\gamma$ is Euler's constant.
The dispersion relations $\omega_T(k)$ and $\omega_L(k)$ are the solutions
to~(\ref{omega-T}) and~(\ref{omega-L}). The angles 
$\phi_T(\omega,k)$
and $\phi_L(\omega,k)$ satisfy~(\ref{theta-T}) 
and~(\ref{theta-L}).
\section{High-Temperature Expansion}
If the temperature is much greater than the gluon mass parameter $m_g$,
the HTL free energy can be expanded in powers of $m_g/T$.
Such an expansion is contrary to the spirit of HTL perturbation theory, 
which is an expansion in powers of  $g$ and $\delta$ with
$m_g^2$ fixed. It is nevertheless 
interesting, because it can be compared directly with the
perturbative expansion for the QCD free energy.
In this section, we compute the high-temperature expansion for the 
HTL free energy through order $(m_g/T)^4$.
\subsection{Separation of Scales}
The high-temperature expansion for the HTL free energy could be deduced
directly from the final expression~(\ref{total}). However, this is difficult
because of the delicate cancellations between quasiparticle and Landau-damping
terms and between transverse and longitudinal contributions.
It is easier to carry out the high-temperature expansion starting from
dimensionally regularized contour integral expressions for ${\cal F}_T$
and ${\cal F}_L$. If we separate out the free energy of a massless boson
given in~(\ref{freecon}), the transverse free energy~(\ref{transcont})
becomes
\bqa
\label{conlog1}
{\cal F}_{T}&=&-{\pi^2\over90}T^4+{1\over2}\int_C{d\omega\over2\pi i}
\int_{\bf k}
\log\left(1-{\Pi_T(-i\omega,k)\over\omega^2-k^2}\right)
{1\over e^{\beta\omega}-1}.
\eqa
Since the logarithm goes to zero as $\omega\rightarrow\infty$, there is no 
need for a convergence factor $e^{\eta\omega}$. We can choose the logarithmic
branch cut in~(\ref{conlog1}) to run from $-\omega_T(k)$ to $-k$ and
from $+k$ to $+\omega_T(k)$. The contour $C$ can then be deformed so that
it wraps around the branch cuts  that run along the real axis between
$-\omega_T(k)$ and $+\omega_T(k)$. In the expression~(\ref{sepsum})
for the longitudinal free energy, the first term can be evaluated 
analytically using~(\ref{logana}). In the second term, it is convenient
to subtract $\log k^2$ from the integrand before writing it as a contour
integral as in~(\ref{locont}). The resulting expression is
\bqa
\label{conlog}
{\cal F}_{L}&=&-{\sqrt{3}\over4\pi}m_g^3T+{1\over2}\int_{\bf k}
\int_C{d\omega\over2\pi i}
\log\left(1-{\Pi_L(-i\omega,k)\over k^2}\right)
{1\over e^{\beta\omega}-1},
\eqa
where
the contour $C$ wraps around
the branch cuts that run along the real axis between $-\omega_L(k)$
and $+\omega_L(k)$.

The integrals in~(\ref{conlog1}) and~(\ref{conlog}) involve two
energy and momentum scales: the ``hard'' scale $T$ and the ``soft''
scale $m_g$. The terms in the high-temperature expansion can receive
contributions from both the hard scale and the soft scale.
Dimensional regularization makes it easy to separate these contributions.
We can obtain the soft contribution by expanding the Bose-Einstein
factor in powers of $\omega/T$:
\begin{equation}
\label{bosexp}
{1 \over e^{\beta \omega} - 1 } = \sum_{n=0}^{\infty} { B_n \over n! } 
\left({\omega\over T}\right)^{n-1} \, ,
\end{equation}
where $B_n$ are the Bernoulli numbers.  
The only scale in the resulting dimensionally regularized integral is $m_g$.
Thus, upon inserting~(\ref{bosexp}) into~(\ref{conlog1}) and~(\ref{conlog}) ,
we obtain expansions in powers of $m_g/T$.
These are the soft contributions to the high-temperature expansion.
We can obtain the hard contributions by expanding the logarithms
in~(\ref{conlog1}) and~(\ref{conlog}) in powers of 
$m_g^2$ or, equivalently, in powers of $\Pi_T$ and $\Pi_L$:
\bqa
\label{expt}
\log\left(1-{\Pi_T(-i\omega,k)\over\omega^2-k^2}\right)&=&-\sum_{n=1}^{\infty}{1\over n}\left[{\Pi_T(-i\omega,k)\over\omega^2-k^2}\right]^n,\\
\label{expl}
\log\left(1-{\Pi_L(-i\omega,k)\over k^2}\right)&=&-\sum_{n=1}^{\infty}{1\over n}\left[{\Pi_L(-i\omega,k)\over k^2}\right]^n.
\eqa
The resulting dimensionally regularized integrals involve only the scale $T$.
Thus, upon inserting~(\ref{expt}) and~(\ref{expl})
into~(\ref{conlog1}) and~(\ref{conlog}), we obtain expansions in powers
of $m_g^2/T^2$.
These are the hard contributions to the high-temperature expansion.
In the next subsections, we calculate the expansions for the soft and the
hard contributions through order $(m_g/T)^4$.
\subsection{Soft Contributions to ${\cal F}_T$ and ${\cal F}_L$}
The soft contributions to the high-temperature expansion
of the free energy consist of the first term
in~(\ref{conlog}) and the soft contributions from the second terms in~(\ref{conlog1})
and~(\ref{conlog}).
We first consider the soft contribution to the transverse free energy.
After inserting the expansion~(\ref{bosexp}) into~(\ref{conlog1}),
there are no longer any 
poles on the imaginary $\omega$ axis.  
Having chosen the branch cuts to lie on the real axis in the interval
$-\omega_T < \omega < \omega_T$, the contour $C$ 
can be taken to infinity in all directions.
Making the change of variables $z=1/\omega$, 
the contour becomes a counterclockwise circle
around $z=0$. The soft contribution to ${\cal F}_T$
can then be written as 
\begin{equation}
	- { 1 \over 2 } 
		\sum_{n=0}^{\infty} { B_n \over n! }T^{1-n}
		\int_{\bf k}
		\oint { dz\over 2 \pi i }z^{-n-1}  
		\log\left( 1 - {\Pi_T(-i/z,k) \over(1/z)^2 - k^2 } \right)
		\, .
\label{softeq1}
\end{equation}
The logarithmic factor is an even function of $z$ that is analytic at $z=0$.
It can therefore be expanded as a power series  in $z^2$:

\begin{equation}
\log\left( 1 - {\Pi_T(-i/z,k) \over (1/z)^2 - k^2 } \right) = \sum_{p=1}^{\infty} T_p(m,k) z^{2p} \, ,
\end{equation}
where $T_p(m,k)$ is a polynomial of degree $p$ in $m^2$ and $k^2$.
Inserting
this expansion into~(\ref{softeq1}), the contour integral vanishes unless
$n=2p$. 
The soft contribution~(\ref{softeq1}) then reduces to
\begin{equation}
	- { 1 \over 2 } 
		\sum_{p=1}^{\infty} { B_{2p} \over (2p)! } T^{1-2p}
		\int_{\bf k}T_p(m,k)
		\, .
\label{softeq2}
\end{equation}
However, since the integrand $T_p(m,k)$ is a polynomial in $k,$
the dimensionally regularized integral over ${\bf k}$ vanishes. 
The soft contribution to ${\cal F}_T$ is therefore zero.

We next consider the soft contributions to the longitudinal free energy 
coming from the second term in~(\ref{conlog}).
Performing the change of variables $z=1/\omega$ yields
\begin{equation}
	- { 1 \over 2 }
		\sum_{n=0}^{\infty} { B_n \over n! } T^{1-n}
		\int_{\bf k}
		\oint{ dz \over 2 \pi i  }z^{-n-1}  
		\log\left( 1 - {\Pi_L(-i/z,k) \over k^2 } \right),
\label{softeq3}
\end{equation}
where the contour is a counterclockwise circle around $z=0$.
The logarithmic factor is an even function of $z$ that it is
analytic at $z=0$. It therefore can be expanded as a power series in
$z^2$:

\begin{equation}
\log\left( 1 - {\Pi_L(-i/z,k) \over k^2 } \right) = \sum_{p=1}^{\infty} L_p(m,k) z^{2p} \, ,
\end{equation}
where $L_p(m,k)$
is a polynomial of degree $p$ in $m^2$ and $k^2$.  
The contour integral in~(\ref{softeq3}) vanishes unless $n=2p$.
But then the integral over ${\bf k}$ vanishes in dimensional regularization
because $L_p$ is a polynomial in $k$.
Thus the soft contribution to the second term of~(\ref{conlog})
is zero.
The only contribution to the free energy from the soft momentum scale
$m_g$ is therefore the first term in~(\ref{conlog}), which comes from the
$n=0$ Matsubara mode.
\subsection{Hard Contribution to ${\cal F}_T$}
The hard contributions to the high-temperature expansion of ${\cal F}_T$
consist of the $T^4$ term in~(\ref{conlog1})
and the power series in $m_g^2$ obtained by inserting~(\ref{expt}) 
into~(\ref{conlog1}).
The first term in that expansion is 
\bqa
\label{h1}
{\cal F}_T^{(1)}=-{1\over2}
\int_{\bf k}
\int_C{d\omega\over2\pi i}
{\Pi_T(-i\omega,k)\over\omega^2-k^2}
{1\over e^{\beta\omega}-1}.
\eqa
The function multiplying the Bose-Einstein factor can be written as
\bqa
\label{multi}
{\Pi_T(-i\omega,k)\over\omega^2-k^2}=
{3m_g^2\over2k^2}\left({\omega^2\over\omega^2-k^2}-{\omega\over2k}
\log{\omega+k\over\omega-k}\right).
\eqa
Thus the integrand in~(\ref{h1}) is the sum of two terms, one
with poles at $\omega=\pm k$ and the other with a branch cut that runs from 
$\omega=-k$ to $\omega=k$.
The integral of the pole terms can be evaluated using the residue theorem.
In the branch cut term, the contour can be collapsed onto the branch cut, and
then
expressed as an integral over positive values of $\omega$.
The resulting expression is 
\bqa 
\label{f1pol}
{\cal F}_{T}^{(1)}
&=&{3\over4}m_g^2
\int_{\bf k}\left[
{1\over k}\left(
{1\over e^{\beta k}-1}+{1\over2}\right)-{1\over k^3}\int_{0}^{k}
d\omega\;\omega\left({1\over e^{\beta\omega}-1}+{1\over2}\right)\right].
\eqa
The integrals can be evaluated analytically. In the last term, the integrand
is evaluated by integrating first over $k$ and then over $\omega$.
The final result is
\bqa
\label{f1cut}
{\cal F}_{T}^{(1)}
&=&-{\pi^2\over16}m^2_gT^2\left({T^2\over\mu^2}\right)^{-\epsilon}
{\Omega_d\over(2\pi)^d}
\left[{1\over\epsilon}+2\log{\zeta^{\prime}(-1)\over\zeta(-1)}-2\log(2\pi)
-2\right].
\eqa
The second term in the expansion of
${\cal F}_T$ in powers of $m_g^2$ is
\bqa
\label{n=2}
{\cal F}_T^{(2)}=-{1\over4}
\int_{\bf k}\int_C{d\omega\over2\pi i}
\left({\Pi_T(-i\omega,k)\over\omega^2-k^2}\right)^2
{1\over e^{\beta\omega}-1}.
\eqa
Using~(\ref{multi}), we can 
separate the integrand from~(\ref{n=2}) into a double
pole term, a double logarithm term, 
and terms that are products of a single pole and
a logarithm.
After using the residue theorem and collapsing the contour onto the 
branch cut, the expression~(\ref{n=2}) reduces to
\bqa\nonumber
{\cal F}_T^{(2)}&=&{9\over32}m_g^4
\int_{\bf k}
\left\{
{3+2\epsilon\over k^3}\left({1\over e^{\beta k}-1}+{1\over2}\right)\right.
+{2\over k^6}\int_0^kd
\omega\;
\omega^3{\partial\over\partial\omega}\left({1\over e^{\beta\omega}-1}\right)
\log{k+\omega\over k-\omega}
\\
&&\left.
-{8\over k^6}\int_{0}^{k}d\omega\;\omega^2
\log{k+\omega\over k-\omega}\left({1\over e^{\beta\omega}-1}
+{1\over2}\right)\right\}.
\eqa
The double integrals can be evaluated by integrating first over 
$k$ and then over $\omega$.
Expanding around
$\epsilon =0$ and keeping terms only through
order $\epsilon^0$, the final result is
\bqa
{\cal F}_T^{(2)}&=&-{3\over16}m_g^4
\left({T^2\over\mu^2}\right)^{-\epsilon}{\Omega_d\over(2\pi)^d}
\left[{8\log2-5\over8}\left({1\over\epsilon}-2\log(2\pi)+2\gamma+{1\over3}\right)-{7\over8}+{\pi^2\over12}\right. 
\label{T2}
\Bigg].
\eqa
\subsection{Hard contribution to ${\cal F}_L$}
The hard contributions to the high-temperature expansion of  
${\cal F}_L$ can be obtained by inserting the
expansion~(\ref{expl}) into~(\ref{conlog}). 
The first term in that expansion is
\bqa
{\cal F}_{L}^{(1)}=-{1\over2}
\int_{\bf k}
\int_C{d\omega\over2\pi i}
{\Pi_L(-i\omega,k)\over k^2}{1\over e^{\beta\omega}-1}.
\eqa
Inside the contour $C$, the integrand has a pole at $\omega=0$ from the 
Bose-Einstein factor and a branch cut in $\omega$ running
from $-k$ to $k$ from the $\Pi_L$ factor.
The residue of the pole is $-3m_g^2/k^2$. Since there is no scale in the
subsequent integral over momentum, the integral vanishes
using dimensional regularization.
The pole can therefore be ignored.
Collapsing the contour onto the branch cut, the expression reduces to
\bqa
{\cal F}_{L}^{(1)}&=&{3\over2}m_g^2{\Omega_d\over(2\pi)^d}
\int_0^{\infty}dk\;k^{-1-2\epsilon}\int_{0}^kd\omega\;
\omega\left({1\over e^{\beta\omega}-1}+{1\over2}\right).
\eqa
Integrating first over $k$ and then over $\omega$, our final result is
\bqa
\label{L1}
{\cal F}_L^{(1)}&=&
{\pi^2\over8}m^2_gT^2\left({T^2\over\mu^2}\right)^{-\epsilon}
{\Omega_d\over(2\pi)^d}
\left[{1\over\epsilon}+2\log{\zeta^{\prime}(-1)\over\zeta(-1)}-2\log(2\pi)\right].
\eqa
Multiplying the corresponding transverse term~(\ref{f1cut}) 
by $2-2\epsilon$ and adding to~(\ref{L1}),
we see that the $1/\epsilon$ poles cancel:
\bqa
\label{totone}
(d-1){\cal F}_{T}^{(1)}+{\cal F}_{L}^{(1)}
={3\over16}m_g^2T^2.
\eqa
The second term in the expansion of ${\cal F}_L$
in powers of $m_g^2$ is
\bqa
\label{lhard}
{\cal F}_{L}^{(2)}&=&-{1\over4}
\int_{\bf k}
\int_C{d\omega\over2\pi i}
\left({\Pi_L(-i\omega,k)\over k^2}\right)^2{1\over e^{\beta\omega}-1}.
\eqa
Inside the contour $C$, the integrand 
has a pole at $\omega=0$ and a branch cut in $\omega$ 
running from $-k$ to $k$. Again the contribution
from the pole vanishes after integrating over $k$ using  
dimensional regularization.
The contour can therefore be collapsed onto the branch cut and~(\ref{lhard})
reduces to 
\bqa
\label{L2f}
{\cal F}_{L}^{(2)}&=&{9\over2}m_g^4
\int_{\bf k}{1\over k^5}
\int_{0}^kd\omega\;
\omega
\left[{\omega\over2k}\log{k+\omega\over k-\omega}-1\right]
\left({1\over e^{\beta\omega}-1}+{1\over2}\right).
\eqa
This can be evaluated by first integrating over $k$ and then over $\omega$.
Expanding around $\epsilon=0$ and including terms through order
$\epsilon^0$, this reduces to 
\bqa
\label{L2}
{\cal F}_{L}^{(2)}&=&-{3\over8}m_g^4
\left({T^2\over\mu^2}\right)^{-\epsilon}
{\Omega_d\over(2\pi)^d}
\left[
(1-\log2)\left({1\over\epsilon}-2\log(2\pi)+2\gamma-{2\over3}\right)\right.
-{\pi^2\over12}\Bigg].
\eqa
Multiplying the corresponding transverse term~(\ref{T2})
by $2-2\epsilon$ and adding to~(\ref{L2}), we obtain
\bqa
\label{f2lf2t}
(d-1){\cal F}_{T}^{(2)}+{\cal F}_{L}^{(2)}
=-{9\over32}m_g^4\left({T^2\over\mu^2}\right)^{-\epsilon}
{\Omega_d\over(2\pi)^d}\left[
{1\over\epsilon}-3+2\gamma-2\log(2\pi)\right].
\eqa
\subsection{High-temperature expansion for ${\cal F}_{\rm HTL}$}
The high-temperature expansion for the HTL free energy is obtained
by inserting the high-temperature expansions for ${\cal F}_T$ and ${\cal F}_L$
into~(\ref{freedef}) along with the counterterm~(\ref{df}).
The expansion includes the $T^4$ term from~(\ref{conlog1}),
the $m_g^3T$ term from~(\ref{conlog}),
the $m_g^2T^2$ term from~(\ref{totone}), and the $m_g^4$ term 
from~(\ref{f2lf2t}).
Combining all the terms, the renormalized free energy is
\begin{eqnarray}
{\cal F}_{\rm HTL} &=& {\cal F}_{\rm ideal} 
\Bigg[ 1 \;-\; {45 \over 4} \left( {3 m_g^2 \over 4 \pi^2 T^2} \right) 
\;+\; 30  \left( {3 m_g^2 \over 4 \pi^2 T^2} \right)^{3/2}
\nonumber
\\
&& \qquad \qquad \qquad
\;+\;  {45 \over 8} \left( \log {\mu_3^2 \over 4 \pi^2 T^2} - 1.232 \right)
	\left( {3 m_g^2 \over 4 \pi^2 T^2} \right)^2 
\;+\; {\cal O} (m_g^6 / T^6) \Bigg].
\label{FHTL-high}
\end{eqnarray}
In the high-temperature limit, the thermal gluon mass parameter approaches
$m_g^2={4\pi\over3}\alpha_sT^2$. 
If we make this identification, the HTL free energy is a selective
resummation of terms in the QCD free energy that are higher orders in
$\alpha_s$.
The expansion parameter $3m_g^2/4\pi^2T^2$
in~(\ref{FHTL-high}) coincides with $\alpha_s/\pi$. 
Thus the high-temperature 
expansion~(\ref{FHTL-high}) can be compared directly with the 
expansion of the QCD free energy in powers of $\alpha_s^{1/2}$
which is given in~(\ref{F1-QCD}).
Comparing the coefficients of the expansions~
(\ref{FHTL-high}) and~(\ref{F1-QCD}), we make the following
observations:
\begin{itemize}
\item{The HTL free energy overincludes the $\alpha_s$ correction by a factor
of three. In conventional perturbation theory, the correction 
$-{15\over 4}\alpha_s/\pi$ arises from interactions between massless
transverse gluons. In HTL perturbation theory, this correction is separated
into two terms: $-{45\over4}\alpha_s/\pi$ from the masses of the
quasiparticles, which is included at leading order, and 
$+{30\over4}\alpha_s/\pi$ from interactions between quasiparticles, which
is included at next-to-leading order.  An explicit calculation of the
next-to-leading order correction is in progress~\cite{twoloophtl}. }
\item{The $\alpha^{3/2}_s$ term, which arises from the effects of electric
screening, is included exactly at leading order in HTL perturbation theory,
provided that the thermal mass parameter is chosen to be 
$m_g^2 = 4 \pi \alpha_s T^2$ up to corrections of order $\alpha_s^2$.}
\item{If we choose the renormalization scale $\mu_3$ in~(\ref{FHTL-high})
to be of order $m_g$, the HTL free energy includes a fraction 
${1\over12}$ of the $\alpha_s^2\log\alpha_s$ correction.}
\end{itemize}
In the HTL free energy,
the oversubtracted $\alpha_s$ term combines with the terms of order
$\alpha_s^{2}$ and higher to give an overall 
correction that is negative in spite of the 
large positive correction from the $\alpha_s^{3/2}$ term.
\section{low-temperature Limit}
In this section, we derive the low-temperature limit
of the HTL free energy for fixed $m_g$. QCD undergoes a phase transition
to a confining phase as the temperature decreases below some critical 
temperature $T_c$. Because of this phase transition, we do not expect 
${\cal F}_{\rm HTL}$ in the limit $T\ll m_g$ to bear any resemblance to the
free energy of QCD at $T<T_c$. However, if ${\cal F}_{\rm HTL}$ is to be used
as a phenomenological model for ${\cal F}_{\rm QCD}$ for $T>T_c$, it is
worthwhile to determine the qualitative behavior of ${\cal F}_{\rm HTL}$
for extreme values of its parameters.

In the low-temperature limit, ${\cal F}_{\rm HTL}$ is
proportional to $m_g^4$. 
The coefficient could be extracted
directly from our final expression~(\ref{total}), but
it is simpler to compute it 
directly from our original expression for the transverse and longitudinal
free energies.
\subsection{Transverse Free Energy}
Our original expression~(\ref{FT-def}) for the transverse free energy
involves a sum over the discrete Matsubara frequencies $\omega_n=2\pi nT$.
As $T\rightarrow 0$, the sum approaches an integral over the continuous
euclidean energy $\omega$:
\bqa
{\cal F}_T\longrightarrow {1\over4\pi}\int_{-\infty}^{\infty}d\omega
\int_{\bf k}\log\left[k^2+\omega^2+\Pi_T(\omega,k)\right].
\eqa
Since $\Pi_T$ is a function of the 
combination $\omega/k$ only, it is convenient
to rescale the energy variable by $\omega\rightarrow k\omega$.
Integrating over the angles of ${\bf k}$ and using the fact that the
integrand is an even function of $\omega$, the integral reduces to
\bqa
{\cal F}_T={1\over2\pi}{\Omega_d\over(2\pi)^d}\mu^{3-d}
\int_0^{\infty}d\omega\int_0^{\infty}dk\;k^d\log\left[(1+\omega^2)k^2+
\Pi_T(\omega,1)\right].
\eqa
The dimensionally regularized integral over $k$ can be evaluated analytically
using~(\ref{logana}),
giving
\bqa
{\cal F}_T={1\over2\pi}{\Omega_d\over(2\pi)^d}
{\Gamma({d+1\over2})\Gamma({1-d\over2})\over d+1}\mu^{3-d}
\int_0^{\infty}d\omega\left[{\Pi_T(\omega,1)\over1+\omega^2}\right]^{{d+1\over2}}.
\eqa
Expanding around $d=3$, we get
\bqa
\label{lowt}
{\cal F}_T&=&-{9\over32\pi}m_g^4
\left({3m_g^2\over2\mu^2}\right)^{-\epsilon}
{\Omega_d\over(2\pi)^d}
\left\{\left({1\over\epsilon}+{1\over2}\right)
\int_0^{\infty}d\omega\;f_T^2(\omega)
-\int_0^{\infty}d\omega\;f_T^2(\omega)\log f_T(\omega)
\right\},
\eqa
where the function in the integrand is
\bqa
f_T(\omega)=\omega\left({\pi\over2}-\arctan \omega\right)-{\omega^2\over1+\omega^2}.
\eqa
The first integral in~(\ref{lowt}) can be evaluated analytically, but the 
second integral must be evaluated numerically. The final result is
\bqa
\label{totalt}
{\cal F}_T=-{9\over32\pi}
m_g^4
\left({3m_g^2\over2\mu^2}\right)^{-\epsilon}
{\Omega_d\over(2\pi)^d}
\left[{\pi(8\log2-5)\over12}\left({1\over\epsilon}+{1\over2}\right)
+0.200871\right].
\eqa
\subsection{Longitudinal Free Energy}
In our original expression~(\ref{FL-def}) for the longitudinal free energy,
the sum over the Matsubara frequencies $\omega_n$ approaches an integral
over $\omega$ in the limit $T\rightarrow 0$. Integrating over the
angles of ${\bf k}$ and rescaling the energy variable by 
$\omega\rightarrow k\omega$, the expression reduces to
\bqa
{\cal F}_L={1\over2\pi}{\Omega_d\over(2\pi)^d}\mu^{3-d}
\int_0^{\infty}d\omega\int_0^{\infty}dk\;k^d\log\left[k^2-
\Pi_L(\omega,1)\right].
\eqa
The integral over $k$ can be evaluated analytically
using~(\ref{logana}). Expanding around
$d=3$, our expression reduces to
\bqa
\label{lowl}
{\cal F}_L=-{9\over8\pi}m_g^4
\left({3m_g^2\over\mu^2}\right)^{-\epsilon}
{\Omega_d\over(2\pi)^d}
\left\{\left({1\over\epsilon}+{1\over2}\right)
\int_0^{\infty}d\omega\;f_L^2(\omega)-\int_0^{\infty}d\omega\;
f_L^2(\omega)\log f_L(\omega)\right\},
\eqa
where the function in the integrand is
\bqa
f_L(\omega)=1+\omega\left(\arctan \omega-{\pi\over2}\right).
\eqa
Evaluating the first integral in~(\ref{lowl}) analytically and the second
numerically, we obtain:
\bqa
\label{totall}
{\cal F}_L=-{9\over8\pi}m_g^4
\left({3m_g^2\over\mu^2}\right)^{-\epsilon}
{\Omega_d\over(2\pi)^d}
\left[{\pi(1-\log2)\over3}\left({1\over\epsilon}+{1\over2}\right)+0.176945\right].
\eqa
\subsection{Renormalized Free Energy}
We can obtain the $T\rightarrow 0$ limit of the HTL free energy by
inserting~(\ref{totalt}),~(\ref{totall}), and the counterterm~(\ref{df}) 
into the expression~(\ref{freedef}) for ${\cal F}_{\rm HTL}$.
The result is
\begin{equation}
{\cal F}_{\rm HTL} \;\to\; (N_c^2 - 1) 
\left[ {9 \over 64 \pi^2} 
	\left( \log{m_g \over \mu_3} - 0.332837 \right)
	m_g^4\right],
\label{FHTL-low}
\end{equation}
which is 
identical to the $m_g^4$ term in our final 
expression~(\ref{total}) for the renormalized free energy.
This is an efficient way of computing the constant under the logarithm
in that term. 

Among the next-to-leading order terms in
the low-temperature expansion for ${\cal F}_{\rm HTL}$ is the 
$m_g^2T^2$ term in~(\ref{total}). To identify the remaining next-to-leading
order terms, we consider the $T\rightarrow 0$ limit of each of the
integrals in~(\ref{total}). The first integal in~(\ref{total}) is the
sum of a transverse term involving $\omega_T$, a longitudinal 
term involving $\omega_L$, and a subtraction term. 
The integral of the subtraction term gives a contribution proportional to
$T^4$, and thus contributes only at next-to-leading order in $T^2/m_g^2$.
The transverse and longitudinal terms are dominated
by the region of small momentum $k$. The dispersion relations in this region
can be approximated by 
\bqa
\label{T-lim}
\omega_T(k)&\longrightarrow&m_g+{3\over5}{k^2\over m_g},\\
\label{L-lim}
\omega_L(k)&\longrightarrow&m_g+{3\over10}{k^2\over m_g}.
\eqa
Making the approximation $\log(1-e^{-\beta\omega})\approx -e^{-\beta\omega}$
in the integrand, we see that both the transverse and the longitudinal
terms reduce to Gaussian integrals in $k$. Their contributions to the 
free energy scale as $T\left(m_gT\right)^{3/2}e^{-m_g/T}$, which falls
faster than any power of $T$.

The Landau-damping term in~(\ref{total}) involves the angles 
$\phi_L$ and $\phi_T$ defined
in~(\ref{theta-T}) and~(\ref{theta-L}). The Bose-Einstein factor
in~(\ref{total}) constrains $\omega$ to be of order $T$. 
In the integral over $k$, there is a logarithmic ultraviolet divergence
proportional to $m_g^2$ that cancels between $\phi_L$ and $\phi_T$.
The corresponding infrared cutoff is of order $m_g$ for the $\phi_L$
integral and of order $(m_g\omega)^{1/2}$ for the $\phi_T$ integral.
Thus the leading contribution from the Landau-damping term is proportional
to $m_g^2T^2\log(m_g/T)$.

The low-temperature limit of ${\cal F}_{\rm HTL} $ is sensitive 
to the value of $\mu_3$.  In particular, 
the coefficient of the $m_g^4$ term in 
(\ref{FHTL-low}) changes sign at $\mu_3^* \approx 0.717 m_g$.
For larger values of $\mu_3$, the ratio 
${\cal F}_{\rm HTL}/{\cal F}_{\rm ideal}$ approaches $+\infty$ 
like $1/T^4$ as 
$T\longrightarrow 0$ with $m_g$ fixed.
For smaller values of $\mu_3$, the ratio approaches $-\infty$ like $1/T^4$.
For $\mu_3=\mu_3^*$, the divergence is less severe 
approaching $-\infty$ like $1/T^2$.
\section{Thermodynamics}
In this section, we examine the thermodynamic functions at leading order
in HTL perturbation theory. 
We first derive a convenient 
expression for the trace anomaly density ${\cal E}-3{\cal P}$
in terms of derivatives  of the parameters $m_g$
and $\mu_3$ with respect to $T$. 
We give a prescription for the temperature dependence 
of $m_g$ which is motivated by the high-temperature limit of
QCD. We then compare our leading order HTL calculations for 
the pressure ${\cal P}$ and for ${\cal E}-3{\cal P}$
with results from lattice simulations of pure-glue
QCD
and with gluon quasiparticle models.

\subsection{Energy Density}
Once the free energy ${\cal F}(T)$ is given as a function of $T$, all 
other thermodynamic functions are determined. In particular, the pressure
${\cal P}$ and the energy density ${\cal E}$ are
\bqa
{\cal P}&=&-{\cal F},\\ 
{\cal E}&=&{\cal F}-T{d{\cal F}\over d T}.
\eqa
The combination ${\cal E}-3{\cal P}$ 
can 
be written as
\bqa
\label{em3p}
{\cal E}-3{\cal P}=-T^5{d \ \over d T}\left({{\cal F}\over T^4}\right).
\eqa
This combination is  proportional to the trace of the energy-momentum
tensor. In QCD with massless quarks, it is nonzero only because scale
invariance is broken by renormalization effects. We will call it the trace
anomaly density. It of course vanishes for an ideal gas of massless particles.
However, it also vanishes for a gas of quasiparticles whose masses are
linear in $T$ and whose interaction are governed by a dimensionless
coupling constant that is independent of $T$.

Our expression~(\ref{total}) for the HTL free energy is a function of
three variables: $T$, $m_g$, and $\mu_3$.
Thus the temperature dependence of ${\cal F}_{\rm HTL}$ is determined
only after the $T$-dependence of $m_g(T)$ and $\mu_3(T)$ is
specified. Using the chain rule, the expression~(\ref{em3p}) for
${\cal E}-3{\cal P}$ can be written as
\bqa
\label{em3p2}
{\cal E}_{\rm HTL}-3{\cal P}_{\rm HTL}=
-T^5\left[{\partial \ 
\over\partial T}+{dm_g\over dT}{\partial \ \ \ \ \over\partial m_g}
+{d\mu_3\over dT}{\partial \ \ \ \over\partial\mu_3}\right]
{{\cal F}_{\rm HTL}\over T^4}.
\eqa
But ${{\cal F}_{\rm HTL}/T^4}$ is a homogeneous function of degree zero in 
the variables $T$, $m_g$, and $\mu_3$, and it therefore satisfies
\bqa
\left[T{\partial  \ \over\partial T}+m_g{\partial \ \ \ \over\partial m_g}
+\mu_3{\partial \ \ \ \over\partial \mu_3}\right]
{{\cal F}_{\rm HTL}\over T^4}=0.
\eqa
This can be used to eliminate the partial derivative with respect to $T$
from~(\ref{em3p2}) :
\bqa
\label{97}
{\cal E}_{\rm HTL}-3{\cal P}_{\rm HTL}=
-T{d \ \over d T}\left(\log{m_g\over T}\right)
m_g{\partial \ \ \over\partial m_g}{\cal F}_{\rm HTL}
-T{d \ \over d  T}\left(\log{\mu_3\over T}\right)
\mu_3{\partial \  \ \over\partial\mu_3}{\cal F}_{\rm HTL}.
\eqa
This expression vanishes if $m_g$ and $\mu_3$ are exactly linear in $T$.
The partial derivative with respect to $\mu_3$ in~(\ref{97}) is simple.
The partial derivative with respect to $m_g$ is more complicated
because in addition to the 
explicit dependence on $m_g$, there is the implicit
dependence of then functions $\omega_T$, $\omega_L$, $\phi_T$, and $\phi_L$
on $m_g$. 
\subsection{$T$-dependence of $m_g$ and $\mu_3$}
Our leading order results for the thermodynamic functions depend on the
thermal gluon mass parameter $m_g$ 
and on a renormalization scale $\mu_3$.
These parameters are completely arbitrary in the sense that the dependence
on them will be systematically canceled by higher orders in HTL perturbation
theory. Since they are being used as a device to reorganize
the perturbative series for thermal quantities, they should depend on 
the temperature $T$.
If higher order calculations in HTL perturbation theory were available,
reasonable values of $m_g(T)$ and $\mu_3(T)$ could be 
determined by the condition that the higher order corrections
be well behaved. In the absence of such calculations, the best we can do is
to make physically motivated estimates of $m_g(T)$ and $\mu_3(T)$.

A useful source of intuition on reasonable values  of $m_g$ is the
high-temperature limit of QCD. In this limit, the expression for the 
thermal gluon  mass is
\bqa
\label{mg}
m_g^2(T)={4\pi N_c\over9}\alpha_s(\mu_4)T^2,
\eqa
where $\mu_4$ is a renormalization scale proportional to $T$.
If we are to use~(\ref{mg}), we must also specify the $T$-dependence of
$\mu_4$. The expression~(\ref{mg}) can be derived in the form of a 
sum-integral over euclidean four-momentum~\cite{Braaten-Nieto}:
\bqa
m_g^2(T)={8\pi N_c\over3}\alpha_s(\mu_4)\sumint_K{{\bf k}^2-k_0^2\over (K^2)^2} ,
\eqa
where $K^2=k_0^2 + {\bf k}^2$.  
The only momentum scales in the sum-integral are integer multiples of 
the lowest Matsubara frequency $2\pi T$.
This suggests that a reasonable value is $\mu_4(T)\approx 2\pi T$.

If we use a parametrization of the running coupling constant $\alpha_s(\mu_4)$
that diverges at some small momentum scale, our
prescription~(\ref{mg}) for the thermal gluon mass will diverge at a 
sufficiently small temperature. 
A parameterization of the running coupling constant
that includes the effects of two-loop running is
\bqa
\label{2lrun}
\alpha_s(\mu_4)={12\pi\over11N_c\bar{L}}\left(
1-{102\over121}{\log\bar{L}\over\bar{L}}\right),
\eqa
where $\bar{L}=\log\left(\mu_4^2/\Lambda^2_{\overline{\mbox{\scriptsize MS}}}\right)$. 
This expression~(\ref{2lrun}) diverges at 
$\mu_4=\Lambda_{\overline{\mbox{\scriptsize MS}}}$, so the thermal gluon mass
$m_g(T)$ in~(\ref{mg}) diverges at the temperature $T$ given by
$\mu_4(T)=\Lambda_{\overline{\mbox{\scriptsize MS}}}$.

In their original paper on screened perturbation theory~\cite{K-P-P},
Karsch, Patk\'os, and Petreczky chose the mass of the scalar quasiparticle
to be the solution of a one-loop gap equation. An analogous choice for
the gluon mass parameter would be to take $m_g(T)$ to be the solution
of the integral equation
\bqa
m_g^2={8N_c\over3\pi}\alpha_s(\mu_4)\int_0^{\infty}
dk\;k{1\over e^{\omega_T/T}-1}.
\eqa
In the limit $\alpha_s(\mu_4)\ll 1$, the solution reduces to~(\ref{mg}),
with the leading correction proportional to $\alpha_s^{3/2}$.  This 
correction would contribute an additional $\alpha_s^{3/2}$ term to 
${\cal F}_{\rm HTL}$ through the $m_g^2$ term in (\ref{FHTL-high}).
It would be canceled at next-to-leading order in HTL perturbation theory
by a similar contribution from the $m_g^2$ term from the one-loop diagram
with an HTL counterterm.  In order to get the correct 
$\alpha_s^{3/2}$ term at leading order in HTL perturbation theory, $m_g^2$
must be given by (\ref{mg}) up to corrections of order $\alpha_s^2$.
We choose here to use the simplest possibility~(\ref{mg}).

We next consider the renormalization scale $\mu_3$. The dependence on 
$\mu_3$ arises because the one-loop HTL free energy has a logarithmic
ultraviolet divergence proportional to $m_g^4$. The divergence has a
three-dimensional origin: it arises as the momentum ${\bf k}\to\infty$
with fixed energy $\omega$. Similar divergences proportional to 
$g^2T^2m_g^2$ and $g^4T^4$ will appear at next-to-leading order 
and next-to-next-to-leading order in HTL
perturbation theory. The infrared cutoffs on the logarithmically divergent
integrals are provided either by the energy $\omega$, which is of 
order $T$, or by the thermal gluon mass $m_g$.
If we use the weak-coupling limit~(\ref{mg})
for $m_g^2$, the divergences proportional to
$m_g^4$, $g^2T^2m_g^2$, and $g^4T^4$
will cancel exactly, leaving logarithms of
$T/m_g$. These logarithms give
the $\alpha_s^2\log\alpha_s$ terms in the weak-coupling 
expansion~(\ref{F1-QCD}) for the QCD free energy.

The high-temperature limit of the HTL free energy is rather insensitive
to the value of $\mu_3$, but the low-temperature behavior is very sensitive. 
With the parameterization~(\ref{2lrun}) for the running coupling constant,
the thermal gluon mass diverges 
as $T$ approaches
the value where $\mu_4(T)=\Lambda_{\overline{\mbox{\scriptsize MS}}}$.
The HTL free energy can therefore be approximated by the
low-temperature expansion given in~(\ref{FHTL-low}).
If we choose $\mu_3$ to be proportional to $m_g$, 
${\cal F}_{\rm HTL}/{\cal F}_{\rm ideal}$
diverges like $m_g^4(T)$, approaching $+\infty$ for $\mu_3>\mu_3^*$
and $-\infty$ for $\mu_3<\mu_3^*$, where $\mu_3^*=0.717m_g$.
If we choose $\mu_3=\mu_3^*$, the HTL
free energy diverges more slowly as $m_g^2(T)$.
We can minimize the 
pathological behavior of ${\cal F}_{HTL}$ at low temperatures
by choosing the
value $\mu_3(T)\approx 0.717m_g(T)$.
\subsection{Comparison with Lattice Gauge Theory}
Lattice gauge theory can be used to calculate the thermodynamic functions of
QCD from first principles, with all errors under control. We can therefore
compare our leading order HTL results with the correct answer provided by
lattice calculations. Boyd et al.~\cite{lattice-0} have calculated 
the thermodynamic functions of a pure $SU(3)$ gauge theory for 
temperatures up to about 5$T_c$. They used Monte Carlo simulations
with high statistics on a lattice as large as $32^3\times 16$ to
calculate plaquette expectation values for about 20 points in $T$
ranging from 0 to about $5T_c$. They extracted the
trace anomaly density ${\cal E}(T)-3{\cal P}(T)$
directly from the plaquette expectation values at 0 and $T$.
The pressure ${\cal P}(T)$ was extracted from the  plaquette expectation 
values at temperatures between 0 and $T$ by computing an integral.
Their final results
were obtained by extrapolating to the continuum limit and were presented
in the form of continuous interpolation curves.
The results of Boyd et al.~\cite{lattice-0} for ${\cal P}(T)$
and ${\cal E}(T)-3{\cal P}(T)$ are shown in Figs.~\ref{pres} and~\ref{diff}.
\begin{figure}[htb]


\hspace{1cm}
\epsfysize=8cm
\centerline{\epsffile{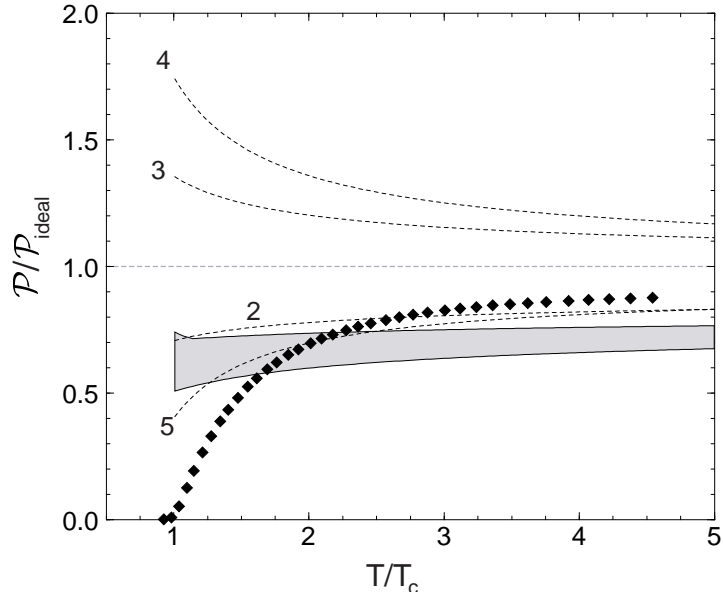}}
\caption[a]{The pressure for pure-glue QCD as a function of
$T/T_c$. The HTL free energy is shown as a shaded band that corresponds to
varying $\mu_3$ and $\mu_4$ by a factor of 2 around their central values.
The weak-coupling expansions through order $\alpha_s$, $\alpha_s^{3/2}$,
$\alpha_s^2$, and $\alpha_s^{5/2}$ are shown as dashed lines
labeled by 
2, 3, 4 and 5. The lattice results of
Boyd et al.~\cite{lattice-0} are shown as diamonds.}
\label{pres}
\end{figure}
 \begin{figure}[htb]


\hspace{1cm}
\epsfysize=8cm
\centerline{\epsffile{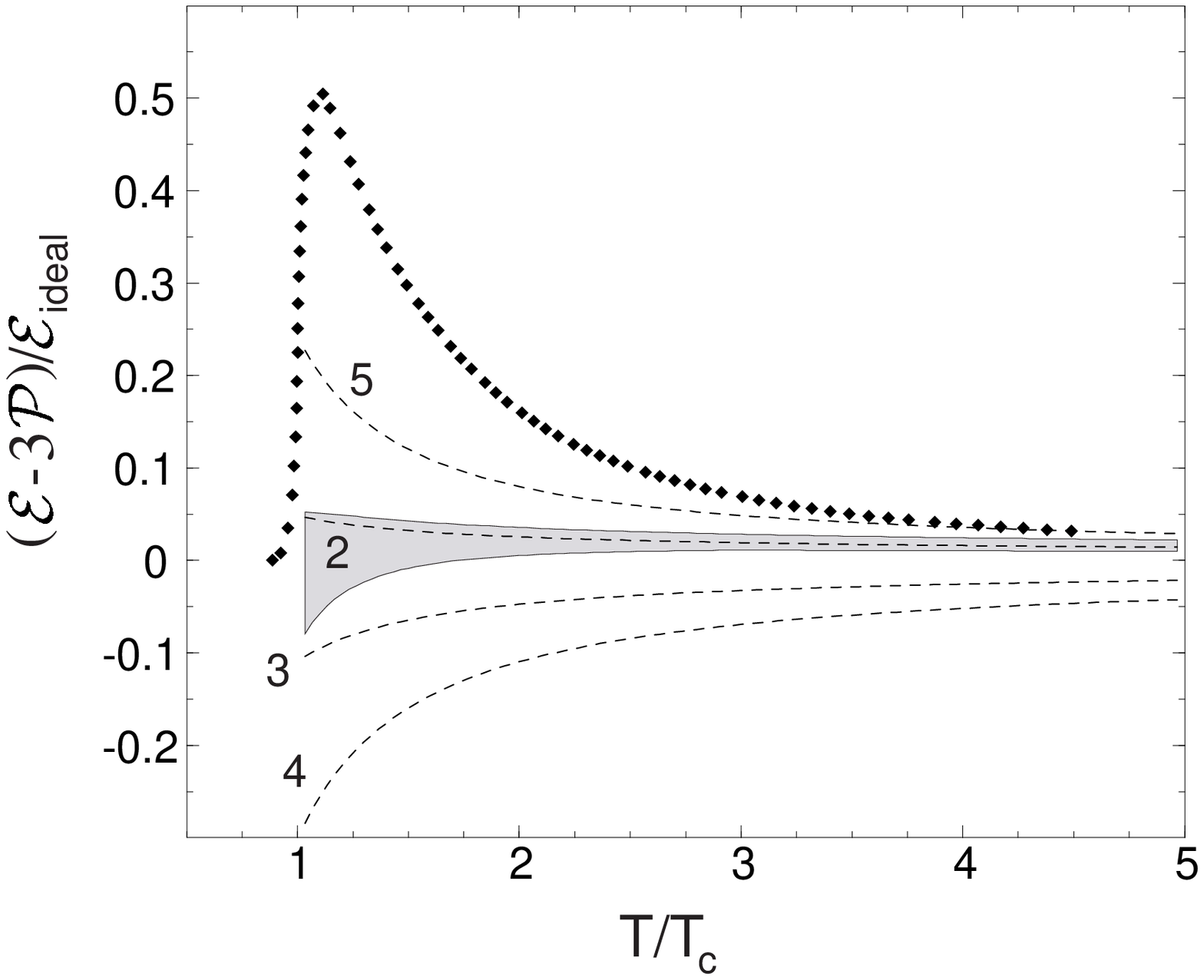}}
\caption[a]{The trace anomaly density for pure-glue QCD as a function of
$T/T_c$.
The HTL trace anomaly density is shown as a
shaded band that correspond to
varying $\mu_3$ and $\mu_4$ by a factor of two around their central values.
The result from differentiating the weak-coupling expansions
through order 
$\alpha_s$, $\alpha_s^{3/2}$,
$\alpha_s^2$, and $\alpha_s^{5/2}$
are shown as dashed lines labeled 2, 3, 4, and 5.
The lattice results ofBoyd et al.~\cite{lattice-0} are shown as diamonds.}
\label{diff}
\end{figure}
The pressure is normalized to the pressure ${\cal P}_{\rm ideal}$
of an ideal gas of massless gluons. The trace anomaly density
is normalized to the energy density
${\cal E}_{\rm ideal}=3{\cal P}_{\rm ideal}$ of the ideal gas.
The normalized energy density ${\cal E}/{\cal E}_{\rm ideal}$ is given
by the sum of the values in Figs.~\ref{pres} and~\ref{diff}.
The interpolation 
curves are represented by 
diamonds at a set of discrete points. The size of the diamonds
indicates the typical error of about 3\% at any single value of $T$.

The leading order HTL results for the pressure and the trace anomaly density
are shown in Figs.~\ref{pres} and~\ref{diff}. We use the expression~(\ref{mg})
for the thermal gluon mass, with the running coupling constant given
by~(\ref{2lrun})
and with $T_c=1.03\Lambda_{\overline{\mbox{\scriptsize MS}}}$.
To illustrate the sensitivity to the choices of 
renormalization scales $\mu_3$ and $\mu_4$, we take their central values
to be 
\bqa
\label{central1}
\mu_3(T)&=&0.717m_g(T),\\
\label{central2}
\mu_4&=&2\pi T,
\eqa
and we allow variations
by a factor of 2 in the coefficients on the right side of~(\ref{central1})
and~~(\ref{central2}). The shaded bands in Figs.~\ref{pres} and~\ref{diff}
indicate the resulting range in the predictions. The ranges are dominated
by the variation in $\mu_4$ at the largest values of $T$ shown and by the
variation in $\mu_3$ for $T$ near $T_c$.
For comparison, the predictions from the QCD weak-coupling
expansion~(\ref{F1-QCD}) with $\mu_4=2\pi T$ are
 also shown in Fig.~\ref{pres} and Fig.~\ref{diff}.
The expansion~(\ref{F1-QCD}) of the pressure
truncated after orders $\alpha_s$, $\alpha_s^{3/2}$, $\alpha_s^2$,
and $\alpha_s^{5/2}$ are shown in Fig.~\ref{pres} as the dashed lines
labeled 2, 3, 4, and 5.
The corresponding predictions for 
the trace anomaly density are shown as dashed lines in Fig.~\ref{diff}.
They were 
obtained by evaluating the derivative in~(\ref{em3p}) 
numerically.

The HTL pressure shown in Fig.~\ref{pres} 
has the correct shape at the largest values of $T$, but the slope also remains
small for $T$ near $T_c$.
The small slope can be understood from the fact that
our expression for $m_g(T)$ in~(\ref{mg}) is almost linear in $T$.
If it were exactly linear in $T$, then ${\cal P}_{\rm HTL}/T^4$
would be constant. The factor of the running coupling constant
in~(\ref{mg}) causes $m_g(T)$ to grow a little slower
than linear in $T$, which causes ${\cal P}_{\rm HTL}/T^4$ to increase slowly
with $T$.

The leading order HTL free energy in Fig.~\ref{pres} lies
below the lattice results at the highest temperature available.
This deviation has the correct sign and
roughly the correct magnitude for the inclusion of the next-to-leading
order correction
in HTL perturbation theory to give a better approximation to the pressure.
The next-to-leading order correction to 
${\cal P}_{HTL}/{\cal P}_{\rm ideal}$ should be positive at large $T$, since
it must approach $+{15\over2}\alpha_s/\pi$ at asymptotically large
temperatures. 

The HTL prediction for the trace anomaly in Fig.~\ref{diff}
is very small. The reason for this is that the differentiation in~(\ref{em3p})
increases the size of the $\alpha_s^{3/2}$ term relative to the
$\alpha_s$ term by a factor of $3/2$. 
As a consequence there is a near cancellation between the oversubtracted
$\alpha_s$ term and the $\alpha_s^{3/2}$ term. If 
the next-to-leading order correction in HTL perturbation
theory
is dominated by the remaining $\alpha_s$ term,
it would give a negative contribution, increasing the
discrepancy with the lattice results. However, the fact that there is a 
near cancellation at leading order makes it dangerous to predict the effect
of the next-to-leading order correction.
\subsection{Comparison with Quasiparticle Models}
The HTL free energy~(\ref{total}) can be interpreted as essentially that
of an ideal gas of gluon quasiparticles, except that it is
modified in such a way as to consistently take into
account the screening effects of the plasma. It is worthwhile to compare
it with the phenomenological 
quasiparticle models that have been used to describe
QCD thermodynamics~\cite{Biro+,Biro2}. In the most recent analyses by 
Peshier et al.~\cite{quasi1} and by L\'evai and
Heinz~\cite{quasi2}, the quasiparticle model is an 
ideal gas of transverse gluons with a temperature-dependent mass.
Their expressions for the pressure and energy density are
\bqa
\label{presquasi}
{\cal P}&=&{N_c^2-1\over3\pi^2}\int_0^{\infty}
dk\;\left(k^3{d\omega_T\over dk}\right){1\over e^{\beta\omega_T}-1}-B(T), \\
{\cal E}&=&
{N_c^2-1\over\pi^2}\int_0^{\infty}
dk\;k^2\omega_T{1\over e^{\beta\omega_T}-1}+B(T).
\eqa
The ``bag function'' $B(T)$, which cancels in the entropy density
${\cal S}=({\cal E}+{\cal P})/T$, is determined up to an integration
constant $B_0$ by the thermodynamic consistency condition
\bqa
{\cal E}=T{d{\cal P}\over dT}-{\cal P}.
\eqa
The transverse dispersion relation $\omega_T$ was taken to be
\bqa
\label{simpledisp}
\omega_T^2=k^2+{3\over2}m_g^2,
\eqa
which is the solution to~(\ref{omega-T}) in the limit $k\gg m_g$. 

We now compare the expression~(\ref{presquasi}) for the pressure in the
quasiparticle model with the HTL pressure which is the negative 
of~(\ref{total}).
The transverse quasiparticle contributions agree after an integration
by parts.
The only difference is the use of the simplified dispersion 
relation~(\ref{simpledisp}) instead of the solution to~(\ref{omega-T}),
but that is a good approximation in the most important momentum regions.
In the quasiparticle model, the contributions from the
plasmon, from Landau damping, and from zero-point energies are
replaced by
the bag function $B(T)$, which
is assumed to cancel in the combination ${\cal E}(T)+{\cal P}(T)$.
This assumption is not justified within the HTL framework.

In their analysis,
L\'evai and Heinz determined the mass $m_g(T)$ by fitting the lattice results
for the pressure ${\cal P}(T)$~\cite{quasi2}. 
Their result for $m_g(T)$ is approximately
linear in $T$ between
$2T_c$ and $4.5T_c$, with a coefficient of proportionality of about $0.75$, 
and it rises to about $3.5T$ as $T$ approaches
$T_c$. This is qualitatively similar to the behavior predicted by the
expression~(\ref{mg}) for the thermal gluon mass in the
weak-coupling limit. If the scale $\mu_4$ in~(\ref{mg})
is determined by fitting the lattice results at the highest available
temperature near 4.5$T_c$, the result is $\mu_4\approx 18T$. This is 
uncomfortably large for the scale of the running coupling constant.
This large value of $\mu_4$ has a simple explanation from the point of view
of HTL perurbation theory. The next-to-leading order contribution to
${\cal P}_{}/{\cal P}_{\rm ideal}$ in HTL perturbation theory will be positive
at large $T$ since it must approach $+{15\over2}\alpha_s/\pi$
in the high-temperature limit. Fitting the lattice results without  allowing
for the next-to-leading order
correction would require a smaller value of $m_g$ and hence
a larger value of $\mu_4$. 
Thus the large value of $\mu_4$ obtained in 
quasiparticle models is an indication that the interaction between the
quasiparticles cannot be neglected.

\section{Outlook}
We have calculated the free energy of a gluon plasma to leading order
in HTL perturbation theory. Extending the calculation to include
quarks is straightforward. A more challenging  problem will be
to extend the calculations to next-to-leading order.
This requires calculating two-loop diagrams in HTL perturbation theory.
Such a calculation
is essential in order to demonstrate that HTL perturbation theory
avoids the convergence problem
that plagues the conventional perturbative 
expansion.

In the high-temperature limit, the two-loop
HTL free energy will agree with the perturbative expansion~(\ref{F1-QCD})
up to errors of order of $\alpha_s^2$. The three-loop HTL free energy
would agree up to order $\alpha_s^3\log\alpha_s$.
In general, the $n$-loop calculation in HTL perturbation theory
will include all the $n$-loop contributions from the scale $T$, which
scale like $(g^2)^{n-1}$, and all the $n$-loop contributions from the
scale $gT$, which scale like $g^{2+n}$.
One limitation of HTL perturbation theory is that it can not 
be used to calculate the contribution from magnetostatic gluons with
momenta of order $g^2T$, which first contribute to the free energy
at order $g^6$. These effects are inherently nonperturbative.
It is possible that they could be calculated by a strategy analogous
to that used by Kajantie et al. to compute the Debye mass for QCD.
The magnetostatic gluons can be described by an effective theory,
called magnetostatic QCD (MQCD), which is a pure $SU(N_c)$ gauge theory
in three euclidean dimensions. 
The parameters of MQCD could perhaps be calculated by HTL perturbation theory,
and then the effects of magnetostatic gluons could be calculated 
nonperturbatively by applying lattice gauge theory methods to MQCD.

In HTL perturbation theory, the leading order quasiparticle dispersion relations
are built into the propagator.  At next-to-leading order in $g$, the quasiparticle
dispersion relations $\omega_T^2$ and $\omega_L^2$ have logarithmic infrared 
divergences proportional to $g^2 T m$ that arise from magnetostatic gluons.
It would be unwise to build the next-to-leading order dispersion relations into the propagator,
because this would introduce a sensitivity to the magnetostatic gluons into
the one-loop free energy that would have to be canceled by higher-loop
diagrams.  The infrared divergences in the next-to-leading order quasiparticle
dispersion relations will of course give a divergent contribution to the 
next-to-leading order free energy through two-loop diagrams that include a
one-loop gluon self-energy correction.  However, we expect this divergence to
be canceled by a divergence in the corresponding Landau-damping contribution.

One of the advantages of HTL perturbation theory is that it can be
applied to the real-time processes that are the most promising signatures
for the quark-gluon plasma. With the exception of the production of hard
dileptons~\cite{ALT-AUR}, these signatures have been calculated only at leading order
in ordinary QCD perturbation theory. There are two reasons to
be concerned about the reliability of these calculations. One is that
the higher order corrections for the signatures are probably at least as
large and unstable as the higher order corrections for the free energy.
There is therefore no way to determine the accuracy of the leading order
calculation.
The other reason for concern is that the conventional weak-coupling expansion
does not give a good approximation to the equation of state. The equation of
state is needed to infer a temperature $T$ from the energy density 
of hot hadronic matter, and $T$ is then used as a parameter in the calculation
of the signatures.
If the method used to calculate the signatures does not reproduce the
equation of state, the whole framework is inconsistent. If the next-to-leading
order calculation in HTL perturbation theory gives a good approximation
to the equation of state, we can have confidence in the predictions for
the signatures that are calculated within the same framework.

HTL perturbation theory provides some justification for quasiparticle models
of the quark-gluon plasma. However, it goes far beyond those models, because
it provides a framework for systematically calculating the effects of
interactions between the quasiparticles.
This could have an enormous impact on the phenomenology of the quark-gluon
plasma, because the physical picture that is suggested by HTL
perturbation theory is dramatically
different from the conventional picture of the quark-gluon plasma 
as an almost ideal
gas of ultrarelativistic quarks and gluons. In HTL perturbation theory,
the quarks and gluons have thermal masses that are comparable to
the temperature $T$, so they are only mildly relativistic. This 
dramatic difference cannot help but have significant phenomenological 
implications.
There have been two previous studies of the effects of quasiparticle
masses on the signatures of a quark-gluon plasma.
Bir\'o, L\'evai and M\"uller~\cite{Biro2} 
have studied the effect of a massive gluon on
the strangeness production in a quark-gluon plasma. They took the $u$ and $d$
quarks to be massless, they neglected the longitudinal mode of the gluon,
and they took the transverse mode to have a temperature-independent mass
of 500 MeV. They found that the gluon mass enhances the production of $s\bar{s}$
pairs at temperatures below 300 MeV.
Additionally, the effects of thermal masses for quarks and gluons on charm production
from a quark-gluon plasma has been studied by L\'evai and Vogt~\cite{levai2}.
They neglected the longitudinal mode of the gluon and took the transverse gluons 
to have the dispersion relation $\omega_T^2=k^2+m_g^2$.
They found that the thermal masses significantly enhanced the thermal
production of charm at RHIC and at LHC. 
 
The previous calculations of the effects of quasiparticle masses have 
serious theoretical inconsistencies, because introducing a transverse gluon
mass by hand destroys gauge invariance. HTL perturbation theory
solves this problem by introducing the transverse gluon mass in a
gauge-invariant way. But the transverse gluon mass is intricately 
linked to other essential aspects of relativistic plasma physics, including
longitudinal gluons, Landau damping, and the screening of interactions.
All of these effects are incorporated consistently  within HTL
perturbation theory. Thus it provides a foundation for developing a
new phenomenology of the quark-gluon plasma in which the many-body
aspects of the system play a central role.

\section*{Acknowledgments}
This work was supported in part by the U.~S. Department of 
Energy Division of High Energy Physics (grant DE-FG02-91-ER40690),
by a Faculty Development Grant 
from the Physics Department of the Ohio State University
and by the National Science Foundation (grant PHY-9800964).

\appendix\bigskip\renewcommand{\theequation}{\thesection.\arabic{equation}}
\setcounter{equation}{0}\section{Integrals}
In this appendix, we collect the results for the integrals
that are required to calculate the one-loop HTL free energy,
the first few terms in the high-temperature expansion, and the low-temperature
limit.
We use dimensional regularization, so that
ultraviolet divergences appear as poles in $\epsilon$.

In the HTL free energy, the ultraviolet divergences are isolated in
subtraction terms
that must be expanded around $\epsilon=0$ through
order $\epsilon^0$.
The integrals required to evaluate the  
quasiparticle subtractions are
\bqa
\label{qpsub1}
\int_0^{\infty}dk\;k^{2-2\epsilon}\sqrt{k^2+m^2}&=&
-{1\over16}m^{4-2\epsilon}\left[{1\over\epsilon}+2\log2-{1\over2}\right],\\
\label{qpsub2}
\int_0^{\infty}dk\;k^{2-2\epsilon}{1\over(k^2+m^2)^{3/2}}&=
&{1\over2}m^{-2\epsilon}\left[{1\over\epsilon}+2\log2-2\right],\\
\label{qpsub3}
\int_0^{\infty}dk\;k^{2-2\epsilon}{1\over(k^2+m^2)^{3/2}}
\log{k^2+m^2\over m^2}&=
&{1\over2}m^{-2\epsilon}\left[{1\over\epsilon^2}-4+{\pi^2\over6}-2\log^22
+4\log2\right].
\eqa
The corresponding integrals required to evaluate the 
Landau-damping subtractions are
\bqa
\label{Ldtr1}
\int_0^{\infty}d\omega\;\omega\int_{\omega}^{\infty}dk\;
k^{-1-2\epsilon}{k^2-\omega^2\over k^2-\omega^2+m^2}&=&-{1\over4}m^{2-2\epsilon}\left[{1\over\epsilon^2}+{\pi^2\over6}\right],\\ 
\label{Ldtr2}
\int_0^{\infty}d\omega\;\omega\int_{\omega}^{\infty}dk\;
k^{-3-2\epsilon}{(k^2-\omega^2)^2\over(k^2-\omega^2+m^2)^2}&=&
{1\over4}m^{-2\epsilon}\left[{1\over\epsilon}-2\right],\\ \nonumber
\int_0^{\infty}d\omega\;\omega^2\int_{\omega}^{\infty}dk\;
k^{-4-2\epsilon}{(k^2-\omega^2)^2\over(k^2-\omega^2+m^2)^2}\log{k+\omega\over k-\omega}&=&{1\over6}m^{-2\epsilon}\left[{1+2\log2\over\epsilon}\right.\\
\label{Ldtr3}
&&\left.+2\log^22-
{22\over3}\log2-{8\over3}\right],\\
\label{Ldlo1}
\int_0^{\infty}d\omega\;\omega\int_{\omega}^{\infty}dk\;
k^{1-2\epsilon}{1\over k^2+m^2}
&=&-{1\over4}m^{2-2\epsilon}\left[{1\over\epsilon}\right],\\
\label{Ldlo2}
\int_0^{\infty}d\omega\;\omega^2\int_{\omega}^{\infty}dk\;k^{-2\epsilon}
{1\over(k^2+m^2)^2}
\log{k+\omega\over k-\omega}&=&{1\over6}m^{-2\epsilon}(1+2\log2)
\left[{1\over\epsilon}-1\right].
\eqa
There is also a $T$-dependent integral that multiplies a pole in $\epsilon$
and must therefore be expanded to order $\epsilon$:
\bqa
\label{Ldtemp1}
\int_0^{\infty}d\omega\;{\omega^{1-2\epsilon}\over e^{\beta\omega}-1}&=
&{\pi^2\over6}T^{2-2\epsilon}\left[
1+\left(2{\zeta^{\prime}(-1)\over\zeta(-1)}-2\log(2\pi)\right)\epsilon
\right].
\eqa

There are several nontrivial  integrals  required to compute the
high-temperature expansion and the low-temperature limit for the 
HTL free energy. One of these integrals is
\bqa
\label{logana}
\int_0^{\infty}dk\;k^{\alpha}\log(k^2+m^2)&=&
{\Gamma({\alpha+1\over2})\Gamma({1-\alpha\over2})\over\alpha+1}m^{\alpha+1}.
\eqa
This integral with $\alpha=2$ is needed to compute the soft contribution
to the free energy from the $n=0$ Matsubara mode. The same integral with
$\alpha=3-2\epsilon$ is needed to compute the low-temperature 
limit of the HTL free energy. In computing the $(m_g/T)^4$
term in the high-temperature expansion, we need the following 
infrared-divergent integral:
\bqa
\label{div}
\int_{0}^{\infty}d\omega\;{\omega^{-1-2\epsilon}\over e^{\beta\omega}-1}
={1\over4}\left({1\over\epsilon}+2\gamma-2\log(2\pi)\right).
\eqa
We also need
the following integral which comes from changing variables in a
momentum integral $x=k/\omega$:
\bqa
\int_1^{\infty}dx\;x^{-4-2\epsilon}\log{x+1\over x-1}
={1+2\log2\over3}+{\pi^2-8\log2-10\over18}\epsilon.
\eqa
Since it 
multiplies an infrared pole in $\epsilon$ from the integral~(\ref{div}),
it must therefore be expanded to order $\epsilon$.

\end{document}